\documentstyle[12pt,fleqn,rotate,epsfig]{article}

\begin{document}

\title{Excitation functions of hadronic observables from SIS to RHIC
energies\thanks{supported by BMBF and GSI Darmstadt}  }

\author{W. Cassing, E. L. Bratkovskaya and S. Juchem \\[5mm]
{\normalsize Institut f\"{u}r Theoretische Physik,}\\
{\normalsize Universit\"{a}t Giessen,}
{\normalsize 35392 Giessen, Germany}    }

\date{ }
\maketitle

\begin{abstract}
We calculate excitation functions for various dynamical quantities
as well as experimental observables from SIS to RHIC energies
within the HSD transport approach which is based on string, quark,
diquark ($q, \bar{q}, qq, \bar{q}\bar{q}$) and hadronic degrees of
freedom without including any explicit phase transition to a
quark-gluon plasma (QGP). It is argued that the failure of this
more 'conventional' approach in comparison to experimental data
should indicate the presence of a different phase which might be
either attributed to space-time regions of vanishing scalar quark
condensate ($<q\bar{q}>$ = 0) or to the presence of a QGP phase
with strongly interacting partons. We study the $K/\pi$ ratio, the
low mass dilepton enhancement in the invariant mass regime from
0.2 -- 1.2 GeV as well as charmonium suppression for central
Au~+~Au collisions as a function of the bombarding energy and
present predictions for these observables as well as hadron
rapiditiy distributions at RHIC energies. Whereas all observables
studied within HSD show smooth increasing/decreasing excitation
functions, the experimental $K^+/\pi^+$ ratio indicates a maximum
at 11 A$\cdot$GeV (or above) which is interpreted as a signature
for a chirally restored phase in the course of the reaction.
\end{abstract}

\vspace{1cm} \noindent PACS: 24.10.-i; 25.75.-q; 11.30.Rd;
13.60.-r


\newpage

\section{Introduction}

The ultimate goal of relativistic nucleus-nucleus collisions is to
reanalyze the early 'big-bang' under laboratory conditions and to
find the 'smoking gun' for a phase transition from the initial
quark-gluon plasma (QGP) to a phase characterized by an
interacting hadron gas \cite{Harris,Shuryak93,QM99}. Any
theoretical approach might describe such a phase transition
starting  from the partonic side with strongly interacting quarks
and gluons \cite{Geiger3,Bernd} or from the hadronic side by
involving hadronic degrees of freedom \cite{Sorge99,Basrev}, i.e.
hadrons with proper self-energies or spectral functions at high
baryon density or temperature. It remains to be seen which
approach will prove to be more successful, economic and
transparent.

Nucleus-nucleus collisions with initial energies per nucleon of
$\sqrt{s}$ = 200~GeV or $\approx$21.5~A$\cdot$TeV will be
available soon at the Relativistic-Heavy-Ion-Collider (RHIC) in
Brookhaven. In central collisions of Au~+~Au here energy densities
above 5 GeV/fm$^3$ are expected such that the critical energy
density for a QGP phase should be overcome in considerable
space-time volumes where the relevant degrees of freedom are
partons (quarks and gluons). Parton cascade calculations have been
used so far \cite{Geiger3,Geiger1,Wang} to estimate the energy
densities and particle production yields in violent reactions at
$\sqrt{s}$ = 200~GeV, an order of magnitude higher than at SPS
energies ($\sqrt{s} \approx$ 20 GeV). Intuitively one expects that
the initial nonequilibrium phase of a nucleus-nucleus collision at
RHIC energies should be described by parton degrees of freedom
whereas hadrons are only formed (by 'condensation') at a later
stage of the reaction which might be a couple of fm/c from the
initial contact of the heavy ions.  Thus parton cascade
calculations -- including transitions rates from perturbative QCD
-- should be adequate for all initial reactions involving a large
4-momentum transfer between the consituents since QCD is well
tested in its short distance properties. The question, however,
remains to which extent the parton calculations can be
extrapolated to low $Q^2$ where hadronic scales become important.
As a rough estimate one can employ here the average mass of vector
mesons, the nucleon and its first excited state, which gives
$Q^2_{crit} \approx$ 1 GeV$^2$. On the other hand, using the
uncertainty relation this implies time scales of $\Delta t<$ 0.2
fm/c = $Q_{crit}^{-1}$ or relative separations of partons $\Delta
r <$ 0.2 fm, which are small compared to the hadronic size or
average life time of the $\rho, \Delta, etc.$ in free space or the
formation time of hadrons $t_F \approx 0.7-0.8$ fm/c as used in
the HSD transport approach \cite{Cass99,Ehehalt,Jgeiss}.

Turning the argument around, a nonequilibrium hadronic approach
involving a time scale of 0.7-0.8 fm/c cannot tell anything about
shorter times because the uncertainty relation does not allow to
distinguish states which are separated in mass by less than
$\approx$ 300 MeV = $t_F^{-1}$, which is the $N-\Delta$ mass
difference.  Thus one faces the problem that neither the parton
description nor a nonequilibrium hadronic model should be valid
for times 0.2 fm/c $\leq t \leq$ 0.7-0.8 fm/c in individual
hadronic reactions, which corresponds to the nonperturbative
formation time of the hadronic wavefunction. This regime of the
'soft' QCD physics is presently not well understood and
appropriate dynamical models are urgently needed. In the HSD
approach these intermediate times are described by color neutral
strings, where the leading quarks and 'diquarks' in a baryonic
string (or quarks and antiquarks in a mesonic string etc.) are
allowed to rescatter again with hadronic cross sections divided by
the number of constituent quarks and antiquarks in the hadrons,
respectively \cite{Jgeiss}.

Furthermore, the question of chiral symmetry restoration at high
baryon density and/or high temperature is of fundamental interest,
too \cite{Harris,Shuryak93}. Whereas lattice QCD calculations at
zero baryon chemical potential indicate that a restoration of
chiral symmetry goes along with the deconfinement phase transition
at some critical temperature $T_c$, the situation is less clear
for finite baryon density where QCD sum rule studies indicate a
linear decrease of the scalar quark condensate $<\bar q q>$ --
which is nonvanishing in the vacuum due to a spontaneous breaking
of chiral symmetry -- with baryon density $\rho_B$ towards a
chiral symmetric phase characterized by $<\bar{q} q>$ = 0. This
decrease of the scalar condensate is expected to lead to a change
of the hadron properties with density and temperature, i.e. in a
chirally restored phase the hadrons might become approximately
massless \cite{Brown} or at least vector and axial vector currents
should become equal \cite{Kochr,Zahed}; the latter implies that
the $\rho$ and $a_1$ spectral functions should become identical.
Since the scalar condensate $<q\bar{q}>$ is not a direct
observable, its manifestations should be found in different
hadronic abundancies and spectra.

Nowadays, our knowledge about the hadron properties at high
temperature or baryon density is based on heavy-ion experiments from 
BEVALAC/SIS to SPS energies where hot and dense nuclear
systems are produced on a timescale of a few fm/c. As mentioned
above, the information from ultrarelativistic nucleus-nucleus
collisions at RHIC, i.e. initial $\sqrt{s} = $ 200 GeV per
nucleon, will be available soon \cite{QM99}. However, any
conclusions about the properties of hadrons in the nuclear
environment are based on the comparison of experimental data with
nonequilibrium kinetic transport theory
\cite{Cass99,Stoecker,Bertsch,Cass90,Koreview}.  As a genuine
feature of transport theories there are two essential ingredients:
i.e. the {\it baryon (and meson) scalar } and {\it vector
self-energies} as well as {\it in-medium elastic} and {\it
inelastic cross sections} for all hadrons involved. Whereas in the
low-energy regime these 'transport coefficients' can be calculated
in the Dirac-Brueckner approach starting from the bare
nucleon-nucleon interaction \cite{Bot,Mal,T3} this is no longer
possible at high baryon density $(\rho_B \geq 2$-$3 \rho_0)$ and
high temperature, since the number of independent hadronic degrees
of freedom increases drastically and the interacting hadronic
system should enter a phase with $<q\bar{q}> \approx$ 0
\cite{Brown,Kochr,GEB,GEB1,Koch} as mentioned before. As a
consequence the hadron self-energies or spectral functions in the
nuclear medium will change substantially especially close to the
chiral phase transition and transport theoretical studies should
include the generic properties of QCD in line with nonperturbative
computations on the lattice \cite{Lat1,Lat2,Lat3,Lat4}.

In this work we concentrate on excitation functions of hadronic
observables from SIS to RHIC energies with the aim to find out
optimal experimental conditions to search for 'traditional'
phenomena such as strangeness enhancement (Section 3), low mass
dilepton enhancement (Section 4) or charmonium suppression in
nucleus-nucleus collisions (Section 5).  Our studies are performed
within the HSD transport approach that has been described in Refs.
\cite{Ehehalt,Jgeiss} and been tested for $p+p$, $p+A$ and $A+A$
collisions from the SIS to SPS energy regime \cite{Cass99}. Actual
predictions for hadron rapidity spectra and $J/\Psi$ suppression
will be presented in Section 6 while Section 7 concludes the study
with a summary.

\section{Theoretical considerations}

In this Section we briefly recall the ingredients of the covariant
transport theory, that has been denoted as {\bf H}adron-{\bf
S}tring-{\bf D}ynamics (HSD) \cite{Ehehalt}, which formally can be
written as a coupled set of transport equations for the
phase-space distributions $f_h (x,p)$ of hadron $h$
\cite{Ehehalt,Cass,Weber1}, i.e.
\begin{eqnarray}
&& \left\{ \left( \Pi_\mu-\Pi_\nu\partial_\mu^p U_h^\nu
   - M_h^*\partial^p_\mu U_h^S \right)\partial_x^\mu
   + \left( \Pi_\nu \partial^x_\mu U^\nu_h+
   M^*_h \partial^x_\mu U^S_h\right) \partial^\mu_p \right\}
   f_h(x,p)\nonumber\\
&& = \sum_{h_2 h_3 h_4\ldots} \int d2 d3 d4 \ldots
   [G^{\dagger}G]_{12\rightarrow 34\ldots}
   \delta^4_{\Gamma}(\Pi +\Pi_2-\Pi_3-\Pi_4 \ldots )  \nonumber\\
&& \times \left\{ f_{h_3}(x,p_3)f_{h_4}(x,p_4)\bar{f}_h(x,p)
   \bar{f}_{h_2}(x,p_2)\right.  \nonumber\\
&& -\left. f_h(x,p)f_{h_2}(x,p_2)\bar{f}_{h_3}(x,p_3)
   \bar{f}_{h_4}(x,p_4) \right\} \ldots\ \ .
\label{Ehg24} \end{eqnarray} In Eq.~(\ref{Ehg24}) $U_h^S(x,p)$ and
$U_h^\mu(x,p)$ denote the real part of the scalar and vector
hadron self-energies, respectively, while $[G^+G]_{12\rightarrow
34\ldots} \delta^4_{\Gamma} (\Pi +\Pi_2-\Pi_3-\Pi_4\ldots )$ is
the 'transition rate' for the process $1+2\rightarrow 3+4+\ldots$
which is taken to be on-shell in the semiclassical limit adopted.
The hadron quasi-particle properties in (\ref{Ehg24}) are defined
via the mass-shell constraint \cite{Weber1},
\begin{equation}
\delta (\Pi_\mu\Pi^\mu-M_h^{*2} ) \ \ ,
\label{Ehg25}\end{equation} with effective masses and momenta (for
a hadron of bare mass $M_h$ and momentum $p^\mu$) given by
\begin{eqnarray}
M_h^* (x,p)&=&M_h + U_h^{{S}}(x,p) \nonumber \\
\Pi^\mu (x,p)&=&p^\mu-U^\mu_h (x,p)\ \ ,
\label{Ehg26}\end{eqnarray}
while the phase-space factors
\begin{equation}
\bar{f}_h (x,p)=1 \pm f_h (x,p) \label{Ehg26a}\end{equation} are
responsible for fermion Pauli-blocking or Bose enhancement,
respectively, depending on the type of hadron in the final/initial
channel. The dots in Eq.~(\ref{Ehg24}) stand for further
contributions to the collision term with more than two hadrons in
the final/initial channels. The transport approach (\ref{Ehg24})
is fully specified by $U_h^S(x,p)$ and $U_h^\mu(x,p)$ $(\mu
=0,1,2,3)$, which determine the mean-field propagation of the
hadrons, and by the transition rates $G^\dagger G\ \delta^4
(\ldots )$ in the collision term, that describe the scattering and
hadron production/absorption rates.

\begin{figure}[t]
\phantom{a}\vspace*{-2.5cm}
\centerline{\psfig{figure=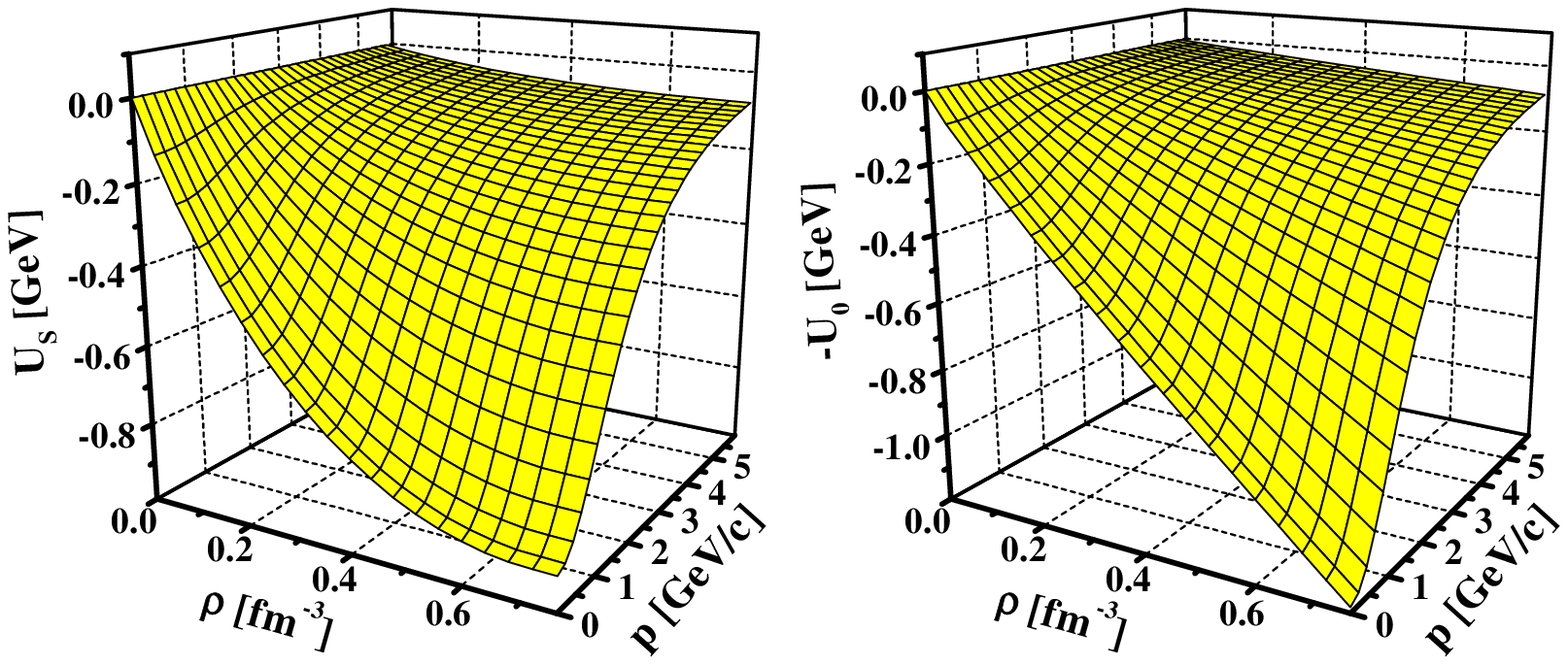,width=16cm}}
\vspace*{-5cm}
\caption{The nucleon scalar ($U_S$) and negative vector potential ($-U_0$)
as a function of the nuclear density $\rho$ and relative momentum
$p$ of the nucleon with respect to the nuclear matter rest frame
as implemented in the HSD transport approach
\protect\cite{Ehehalt}.}
\label{Fig1}
\end{figure}

In Ref. \cite{Ehehalt} the scalar and vector mean-fields $U_h^S$
and $U^\mu_h$ have been determined in the mean-field limit from an
effective hadronic Lagrangian density ${\cal L}_H$ that has been
fitted to the equation of state of nucleonic matter as resulting from 
the Nambu-Jona-Lasinio (NJL) model. Without going through the
arguments again we show in Fig. \ref{Fig1} the nucleon scalar
($U_S$) and negative vector potential ($-U_0$) as a function of
the nuclear density $\rho$ and relative momentum ${\bf p}$ of the
nucleon with respect to the nuclear matter rest frame. Whereas the
vector potential increases practically linearly with density (at
low momentum ${\bf p}$) the scalar potential saturates with
density such that the nucleon effective mass $M^* = M_0+U_S$
almost drops to zero for $\rho \geq$ 0.6 fm$^{-3}$. Both
potentials decrease rather fast in magnitude with momentum ${\bf
p}$ and practically vanish above a few GeV/c.

\begin{figure}[t]
\phantom{a}\vspace*{-3cm}
\centerline{\psfig{figure=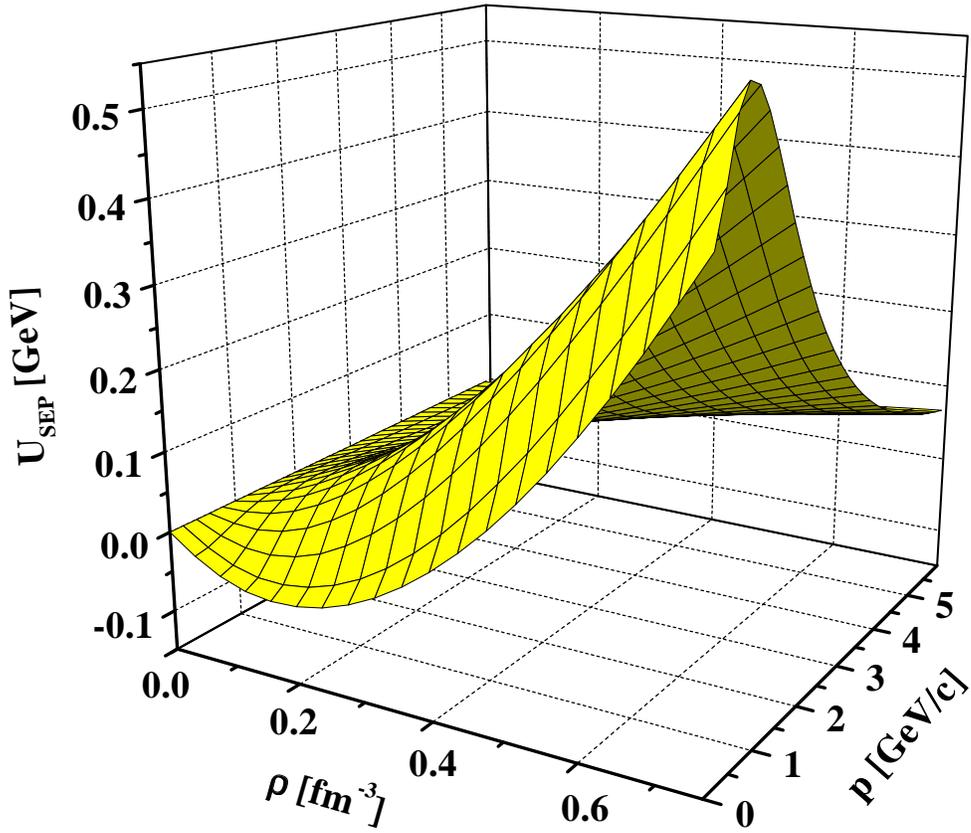,width=16cm}} \vspace*{-5cm}
\caption{The potential $U_{SEP}$ (5) -- as resulting from the
nucleon scalar ($U_S$) and vector potential ($U_0$) in Fig. 1 --
as a function of the nuclear density $\rho$ and relative momentum
$p$ of the nucleon with respect to the nuclear matter rest frame.}
\label{Fig2}
\end{figure}

In Fig. \ref{Fig2} the real part of the potential
\begin{eqnarray}
U_{SEP} &=&  U_0(\rho_0,{\bf p}) + \sqrt{{\bf p}^2 + (M_N +
U_S)^2} - \sqrt{{\bf p}^2 + M_N^2} \label{Ehg23}\end{eqnarray} is
shown again as a function of $\rho$ and ${\bf p}$. Whereas we see
a net attraction for momenta $|{\bf p}| \leq$ 0.5~GeV/c up to
densities of $\approx$ 0.3~fm$^{-3}$, the net potential becomes
repulsive for higher momenta, reaches a maximum repulsion at
$|{\bf p}| \approx$ 1 GeV/c and then drops again with $|{\bf p}|$.
We mention that at density $\rho_0$ the potential $U_{SEP}$
compares well with the potential from the data analysis of Hama et
al. \cite{Hama} as well as Dirac-Brueckner computations from
\cite{Mal} up to a kinetic energy $E_{kin}$ of 1 GeV
\cite{Ehehalt}. The formula (\ref{Ehg23}) reduces to the familiar
expression for the Schroedinger equivalent potential (Eq. (3.16)
of Ref. \cite{Cass99}) in the low density limit.

In our transport calculations we include nucleons, $\Delta$'s,
N$^*$(1440), N$^*$(1535), $\Lambda$, $\Sigma$ and $\Sigma^*$
hyperons, $\Xi$'s and $\Omega$'s  as well as their antiparticles.
In a first approximation we assume here that all baryons (made out
of light $(u,d)$ quarks) have the same scalar and vector
self-energies as the nucleons while the hyperons pick up a factor
2/3 according to the light quark content and $\Xi$'s a factor of
1/3, respectively.

In the HSD approach the high energy inelastic hadron-hadron
collisions are described by the FRITIOF model \cite{LUND}, where
two incoming hadrons emerge the reaction as two excited color
singlet states, i.e. 'strings'. According to the Lund model
\cite{LUND} a string is characterized by the leading constituent
quarks of the incoming hadron and a tube of color flux is supposed
to be formed connecting the rapidly receding string-ends. In the
HSD approach baryonic ($qq-q$) and mesonic ($q-\bar{q}$) strings
are considered with different flavors ($q = u,d,s$). In the
uniform color field of the strings virtual $q\bar{q}$ or
$qq\bar{q}\bar{q}$ pairs are produced causing the tube to fission
and thus to create mesons or baryon-antibaryon pairs. The
production probability $P$ of massive $s\bar{s}$ or
$qq\bar{q}\bar{q}$ pairs is suppressed in comparison to light
flavor production ($u\bar{u}$, $d\bar{d}$) according to a
Schwinger-like formula \cite{Schwinger}
\begin{eqnarray}
{P(s\bar{s}) \over P(u\bar{u})} = \gamma_s = \exp\left(-\pi
{m_s^2-m_q^2\over 2\kappa}, \right)
\label{schwinger}
\end{eqnarray}
with $\kappa\approx 1$~GeV/fm denoting the string tension. Thus in
the Lund string picture the production of strangeness and
baryon-antibaryon pairs is controlled by the constituent quark and
diquark masses. Inserting the constituent quark masses $m_u=0.3$~GeV
and $m_s=0.5$ GeV a value of $\gamma_s \approx 0.3$ is
obtained. While the strangeness production in proton-proton
collisions at SPS energies is reasonably well reproduced with this
value, the strangeness yield for p~+~Be collisions at AGS energies
is underestimated by roughly 30\% (cf. \cite{Jgeiss}). For that
reason the relative factors used in the HSD model are
\cite{Jgeiss}
\begin{eqnarray}
u:d:s:uu = \left\{
\begin{array}{ll}
1:1:0.3:0.07 &, \mbox{at SPS to RHIC energies} \\ 1:1:0.4:0.07 &,
\mbox{at AGS energies} ,
\end{array}
\right. \label{HSD-supp}
\end{eqnarray}
with a linear transition of the strangeness suppression factor
$\gamma_s = s:u$ as
a function of $\sqrt{s}$ in between.

Additionally a fragmentation function $f(x,m_t)$ has to be
specified, which is the probability distribution for hadrons with
transverse mass $m_t$ to acquire the energy-momentum fraction $x$ from 
the fragmenting string,
\begin{eqnarray}
f(x,m_t)\approx {1 \over x} (1-x)^a \exp\left(-bm_t^2/x  \right),
\end{eqnarray}
with $a=0.23$ and $b=0.34$~GeV$^{-2}$ \cite{Jgeiss}.

We recall that the LUND model \cite{LUND} includes partonic
diffractive scattering and mini-jet production as well
\cite{PYTHIA}. The latter phenomena are not important at SPS
energies and below, however, become appreciable at RHIC energies.
In this respect the HSD approach dynamically also includes the
hard partonic processes as far as quarks and antiquarks are
involved. However, it does not employ hard gluon-gluon processes
beyond the level of 'string phenomenology'. This has to be kept in
mind with respect to the predictive power of the model at RHIC
energies and beyond.

The medium modifications due to the hadron self-energies,
furthermore, require to introduce some conserving approximations
in the collision terms in line with the modified quasi-particle
properties. Since these in-medium modifications -- related to 'low
momentum physics' --  are not of primary interest in this study we
discard an explicit discussion here and refer the reader to Ref.
\cite{Ehehalt}.

\subsection{The scalar condensate}

The scalar quark condensate $<q\bar{q}>$ is viewed as an order
parameter for the restoration of chiral symmetry at high baryon
density and temperature. A model independent relation for the
scalar quark condensate at finite (but small) baryon density and
temperature has been given by Drukarev and Levin \cite{Drukarev},
i.e
\begin{equation}
\label{condens} \frac{<q\bar{q}>}{<q\bar{q}>_V} = 1 -
\frac{\rho_S}{f_\pi^2 m_\pi^2} \left[\Sigma_{\pi} + m \frac{d}{dm}
\left(\frac{E(\rho)}{A}\right) \right],
\end{equation}
where $<q\bar{q}>_V$ denotes the vacuum condensate, $\Sigma_\pi \approx$
45 MeV is the pion-nucleon $\Sigma$-term, $f_\pi$ and $m_\pi$ the
pion decay constant and pion mass, respectively, while $E(\rho)/A$ is
the binding energy per nucleon. Since for low
densities the scalar density $\rho_S$ in (\ref{condens}) may be
replaced by the baryon density $\rho_B$, the change in the scalar
quark condensate starts linearly with $\rho_B$ and is reduced by a
factor 1/3 at saturation density $\rho_0$. A simple linear
extrapolation then would indicate that at $\rho_B \approx 3
\rho_0$ a restoration of chiral symmetry might be achieved in
heavy-ion collisions.

A reasonable estimate for the scalar condensate in dynamical
calculations has been suggested by Friman et al. \cite{Toneev98},
\begin{equation}
\frac{<q\bar{q}>}{<q\bar{q}>_V} = 1 - \frac{\Sigma_\pi}{f_\pi^2
m_\pi^2}\rho_S - \sum\limits_h{\sigma_h \rho_S^h \over f_\pi^2
m_\pi^2}, \label{condens2} \end{equation}
where $\sigma_h$ denotes
the $\sigma$-commutator of the relevant mesons $h$. For pions and
mesons made out of light quarks and antiquarks we use $\sigma_h =
m_\pi/2$ whereas for mesons with a strange (antistrange) quark we
adopt $\sigma_h = m_\pi/4$ according to the light quark content.
Within the same spirit the $\Sigma$-term for hyperons is taken as
2/3 $\Sigma_{\pi} \approx$ 30 MeV while for $\Xi$'s we use 1/3
$\Sigma_{\pi} \approx$ 15 MeV.

The scalar density of mesons (of type $h$) is given by
\begin{equation}
\label{mesons} \rho_S^h = \frac{(2s+1) (2t+1)}{(2\pi)^3} \int d^3
{\bf p} \frac{m_h}{\sqrt{{\bf p}^2 + m_h^2}} f_h({\bf r},{\bf
p};t),
\end{equation}
with $f_h$ denoting the meson phase-space distribution
of species $h$. In (\ref{mesons}) $s,t$ denote the spin and
isospin quantum numbers, respectively. We note that the scalar
density of baryons is calculated in line with (\ref{mesons}) by
replacing the mass and momentum by the effective quantities in
(\ref{Ehg26}).

The actual numerical result for the space-time dependence of the
scalar condensate (\ref{condens2}) is shown in Figs. \ref{Fig3}
and \ref{Fig4} for central Au~+~Au collisions at 6~A$\cdot$GeV and
20~A$\cdot$GeV, respectively. Here the condensate is divided by
the vacuum condensate $<q\bar{q}>_V$ such that the nonperturbative
vacuum is characterized by a value of 1. Furthermore, the
$z$-direction has been stretched by the Lorentz-factors
$\gamma_{cm}$ to compensate for Lorentz contraction, while
negative numerical values for the condensate have been suppressed.
According to (\ref{condens}), (\ref{condens2}) the scalar
condensate is reduced inside the approaching nuclei by about 35\%;
this reduction becomes more pronounced when the nuclei achieve
full overlap. As seen from Fig. \ref{Fig3} already at 6
A$\cdot$GeV there is a substantial space-time region of vanishing
scalar condensate for 6~fm/c~$\leq t \leq 13$~fm/c, where the
conventional hadronic picture is not expected to hold anymore. The
space-time volume of vanishing quark condensate slightly increases
for 20 A$\cdot$GeV (Fig. \ref{Fig4}), however, the increase is
only moderate and resembles very much the situation for the
space-time integral of high density baryon matter (cf. Fig. 1 in
Ref. \cite{brat97}).

We mention that at higher bombarding energies the 4-volume ($x =
(t,{\bf r})$)
\begin{eqnarray}
V(\alpha)= \int d^3{\bf r} dt \ \Theta(\alpha -
\frac{<q\bar{q}>_x}{<q\bar{q}>_V}) , \label{volume} \end{eqnarray}
that counts the fraction of the scalar condensate below the value
of $\alpha$ (0 $\leq \alpha \leq$ 0.3) is essentially determined
by the pion density whereas below about 20 A$\cdot$GeV the
dominant contribution stems from the baryons (cf. also Ref.
\cite{Toneev98}). Since the pion density can be considered as a
measure of the vacuum 'temperature', a chiral order transition
below about 20 A$\cdot$GeV is dominated by the baryon density
while especially at SPS or even RHIC energies a chiral order
transition is due to 'temperature'.

The question now arises, if there are proper experimental
observables that allow to trace down such type of phase transition
(or cross over). When gating on central collisions of Au~+~Au (or
Pb~+~Pb) such phenomena should show up in their excitation
functions. We note that in a pure hadronic transport approach we
expect a smooth behaviour of practically all observables with
bombarding energy due to an increase of thermal excitation energy.
This is not so obvious for the HSD approach where a gradual
transition from hadronic excitations to strings and quark (or
diquark) degrees of freedom is involved. In fact, in Ref.
\cite{Sahu99} it has been claimed that the excitation function of
transverse and elliptic proton flow together with the transverse
$p_t$ spectra of protons suggest a transition from {\it hadron} to
{\it string} matter at about 5 A$\cdot$GeV. Here, however, we
concentrate on meson abundancies and spectra and refer the reader
to Ref. \cite{Sahu99} for the collective dynamical aspects and to
Ref. \cite{brat99c} for the thermal properties of the theory that
involves a limiting ('Hagedorn') temperature of $T_S \approx$ 150
MeV due to the continuum string excitations.

\section{Meson production}

To present a general overview on the meson abundancies in
nucleus-nucleus collisions we show in Fig. \ref{Fig5} the meson
multiplicities for central collisions of Au~+~Au from SIS to RHIC
energies.  All multiplicities for $\pi^+, \eta, K^+, K^-, \phi$ as
well as $J/\Psi$ mesons show a monotonic increase with bombarding
energy which is only very steep at 'subthreshold' energies, i.e.
at bombarding energies per nucleon below the threshold in free
space for $NN$ collisions. At higher bombarding energies the meson
abundancies group according to their quark content, i.e. the
multiplicities are reduced (relative to $\pi^+$) by about a factor
of 4--5 for a strange quark, a factor of $\approx$ 50 for
$s\bar{s} \equiv \phi$ and a factor of $\approx 2\cdot 10^{5}$ for
$c\bar{c} \equiv J/\Psi$. We mention that in these calculations
the meson rescattering and reabsorption processes have been taken
into account; this reduces the $J/\Psi$ cross section at RHIC
energies by about a factor of 10 (cf. Section 5).  At
'subthreshold' energies the $\phi$ multiplicity turns out to be
almost the same as the antikaon multiplicity, but then rises less
steeply with bombarding energy. Apart from the $\phi$ excitation
function -- that still has to be controlled experimentally -- we
thus find no pronounced change in the shape of the meson
abundancies up to RHIC energies where data are expected to come up
soon.

We recall that our investigations on strangeness production up to
2~A$\cdot$GeV \cite{Cass97a,Brat97b} have given some evidence for
attractive antikaon potentials in the medium while for kaons only
a very moderate repulsive potential was suggested \cite{Brat97b};
$\eta$ mesons apparently do not show sensible in-medium effects
according to the studies in Ref. \cite{Brat98a,Brat98b} in
comparison to the available experimental spectra. At AGS and SPS
energies, on the other hand, the potential effects on kaon and
antikaon abundancies have been found to be only very low
\cite{Cass99} such that meson potentials have been discarded in
Fig. \ref{Fig5} for the overview.

Since the meson abundancies show no sudden change in the
excitation function, we turn to particle ratios. Here strangeness
enhancement has been suggested for more than 2 decades to possibly
qualify as a sensible probe for a QGP phase (cf. Ref. \cite{SQM98}
for a recent overview). Here we examine the $K^+/\pi^+$ ratio at
midrapidity in central Au~+~Au (or Pb~+~Pb) collisions where
experimental data are now available from SIS to SPS energies
\cite{SQM98,E866,Alard}. We recall that detailed comparisons of
pion and kaon rapidity distributions and transverse momentum
spectra to the available data have been presented in Refs.
\cite{Cass99,Jgeiss} such that we directly can continue with the
corresponding excitation function.

In order to discuss the strangeness production over the complete
energy range from SIS to RHIC energies we show in Fig. \ref{Fig6}
the calculated $K^+/\pi^+$ ratio at midrapidity (-0.5 $\leq y_{cm}
\leq$ 0.5) for central Au~+~Au collisions in comparison to the
experimental data. This ratio experimentally is substantially
lower at SPS energies ($\approx 13.5\%$) compared to AGS energies
($\approx 19\%$). At SPS energies this ratio is only enhanced by a
factor 1.75 for central Pb~+~Pb collisions relative to p~+~p
reactions and has to be compared to the factor $\approx$ 3 at AGS.
Such a decrease of the scaled kaon yield from AGS to SPS energies
is not described by the HSD transport model (without kaon
self-energies) which shows a monotonic increase with bombarding
energy similar to $pp$ reactions (open circles). Furthermore, the
higher temperatures and particle densities at SPS energies tend to
enhance the $K^+/\pi^+$ yield closer to its thermal equilibrium
value of $\approx 20-25\%$ \cite{BM} at chemical freezout and
temperatures of $T\approx 150$ MeV.

Our findings have to be compared to results obtained by other
microscopic approaches. Here only the RQMD model very recently
\cite{Wang99} provides a study partly comparable to that of Ref.
\cite{Jgeiss}, however, excluding p~+~A reactions to control the
amount of $K^+$ production by rescattering. Whereas kaon
production in p~+~p reactions is comparable to our results in
\cite{Jgeiss} (input of the transport model), the RQMD model
yields a higher $K^+/\pi^+$ ratio in Au~+~Au, Pb~+~Pb collisions
at all energies due to a higher kaon yield from rescattering. The
latter can be attributed to 'high mass strange resonances' that
have been incorporated to describe the low energy kaon production
via resonance production and decay \cite{Sorge96} (s-channels).
Since these 'high mass resonances' can be repopulated in
resonance-resonance scattering, a rather high $\sqrt{s}$ is
concentrated in a single degree of freedom for a short time. Such
'hot spots' then lead to a higher $K^+/\pi^+$ ratio in A~+~A
reactions especially at AGS energies. While the $K^+/\pi^+$ ratio
can be reasonably described at AGS energies in Au~+~Au reactions,
this ratio is overestimated significantly at SPS energies (cf.
Fig. 4 of Ref.  \cite{Wang99}). Thus also the RQMD model does not
describe the relative decrease of the $K^+/\pi^+$ ratio from 11
A$\cdot$GeV to 160 A$\cdot$GeV. The results of the RQMD
calculations at SIS energies are not known to the authors.

\section{Dilepton production}

Electromagnetic decays to virtual photons ($e^+ e^-$ or $\mu^+ \mu^-$
pairs) have been suggested long ago to serve as a possible signature
for a phase transition to the QGP
\cite{Shuryak,Kaj,Ruuskanen,Cleym,Uheinz} or to be an ideal probe for
vector meson spectroscopy in the nuclear medium. As pointed out in
Refs. \cite{Zahed,Kling} the isovector current-current correlation
function is proportional to the imaginary part of the $\rho$-meson
propagator and also to the dilepton invariant mass spectra.  Dileptons
are particularly well suited for an investigation of the violent phases
of a high-energy heavy-ion collision because they can leave the
reaction volume essentially undistorted by final-state interactions.
Indeed, the dilepton studies in heavy-ion collisions  by the DLS
Collaboration at the BEVALAC \cite{ro88} and by the
CERES~\cite{CERES,Ullrich}, HELIOS~\cite{HELIOS,HELI2},
NA38~\cite{NA38} and NA50 Collaborations~\cite{NA50} at SPS energies
have found a vivid interest in the nuclear physics community.

We recall that the question of chiral symmetry restoration does
not necessarily imply that vector meson masses have to drop with
baryon density or temperature \cite{KoKoch}. Actually, chiral
symmetry restoration (ChSR) only demands that the isovector
current-current correlation function and the axial vector
current-current correlation function (dominated by the chiral
partner of the $\rho$, the $a_1$-meson) should become identical at
high $\rho_B$ or temperature $T$, respectively, because there
should be no more differences between left- and right-handed
particles or equivalently vector and axial vector currents
\cite{Zahed,KoKoch}.  Thus also a strong broadening of the $\rho$-
as well as the $a_1$-spectral function and their mixing in the
medium \cite{Rapp,friman,RappNPA,Rapp99} can be considered as a
signature for ChSR which, however, is not easy to detect
experimentally.

In Ref. \cite{Cass99} we have demonstrated that the present
experimental data on the low mass dilepton enhancement at SPS
energies can be described equally well within the 'dropping
$\rho$-mass' scenario as well as within the 'melting' $\rho$
picture, which implies a large spreading in mass of the $\rho$
spectral function due to its couplings to baryons and/or mesons.
The situation at SIS/BEVALAC energies is more 'puzzling' since
here the low mass dilepton enhancement is neither described in the
dynamical spectral function approach \cite{Brat98b} nor in the
'dropping mass' scheme \cite{Brat98c}, though the $pp$ dilepton
data from the DLS collaboration \cite{DLSpp} are reproduced within
the known sources rather well from 1 -- 5 GeV bombarding energy
\cite{Brat99a,Ernst}. In short, the dynamical origin of the low
mass dilepton enhancement is not yet understood.

Here we propose to investigate the excitation function of low mass
dilepton enhancement in central Au~+~Au collisions. As discussed in
Refs. \cite{Cass99,Rapp99} this excess of dileptons is most probably
due to the isovector ($\rho$) channel, i.e. the imaginary part of the
isovector current-current correlation function which, however, mixes
with the axial vector current-current correlation function at finite
temperature and baryon density \cite{Zahed,Rapp99}.

Since the $\rho$-meson spectral function is of primary interest,
it is important to have some information about the actual baryon
densities that the $\rho$-meson experiences during its propagation
and decay in central Au~+~Au collisions. This information is
displayed in Fig. \ref{Fig7} -- as resulting from the HSD
transport calculation -- for central reactions at 2, 5, 10, 20, 40
and 160 A$\cdot$GeV. Here the meson-baryon production channels and
baryon-baryon production channels are summed up by the solid lines
and denoted as ($\pi B \rightarrow \rho, BB \rightarrow \rho$).
The meson-meson production channels such as $\pi \pi \rightarrow
\rho, a_1 \rightarrow \pi \rho \ $ etc. are summed up in the
dashed histograms indicated as $\pi \pi \rightarrow \rho$
according to the dominating channel. We find that especially the
initial $BB$ production channels produce $\rho$-mesons at very
high densities close to 2 $\rho_0 \gamma_{cm}$ where $\gamma_{cm}$
is the Lorentz factor in the cms. However, these production
channels are much less abundant than the meson-meson channels
which extend over a larger time span (for the heavy system
Au~+~Au) and essentially occur at much lower baryon density,
respectively. This effect becomes even more pronounced with
increasing bombarding energy, i.e. 40 -- 160 A$\cdot$GeV, where
most of the $\rho$-mesons are produced at baryon densities below 2
$\rho_0$. Since in Fig. \ref{Fig7} the relative abundancy of
$\rho$-mesons is displayed as a function of the baryon density at
the production (formation) point, the time averaged value of the
density is even lower due to a fast expansion of the hadronic
fireball. Thus to probe on average high baryon densities by
$\rho$-mesons one should step down in energy to 2 -- 5 A$\cdot$GeV
in order to optimize effects due to the coupling to baryons.

A general overview of dilepton mass spectra in central ($b$=2 fm)
Au~+~Au collisions from 2 A$\cdot$GeV to 21.5 A$\cdot$TeV is
presented in Fig.  \ref{Fig8}, where the upper part corresponds to
the case of vacuum spectral functions for all mesons (cf. Fig.
8.20 of Ref. \cite{Cass99}), while the lower part is calculated by
employing the $\rho$ spectral function from Rapp et al.
\cite{RappNPA,CasRap}. Whereas for the free meson spectral
function one observes essentially an increase of the dilepton
production channels with bombarding energy without any substantial
change in the spectral shape (except for an increasing peak from
the $\phi$; cf. Fig. 5), the in-medium calculations yield almost
exponential mass spectra above $M \geq$ 0.4 GeV with small peaks from 
vacuum $\omega$ and $\phi$ decays at $M \approx 0.78$ GeV and 1.02~GeV.

We mention that for the vacuum spectral functions (upper part of
Fig. 8) the shape of the dilepton spectra (for $M_{e^+e^-} \geq$
0.15 GeV) is due to the superposition of $\eta, \eta', \omega$ and
$a_1$ Dalitz decays and direct vector meson decays ($\rho, \omega,
\phi$), where all mesons may also be produced in meson-baryon and
meson-meson collisions. The increase in dilepton yield (for
$M_{e^+e^-} \geq $ 0.15 GeV) with bombarding energy thus is due to
an enhanced production of $\eta, \eta', \rho, \omega, \phi, a_1$
mesons. Their relative abundance from SIS to RHIC energies does
not scale directly with the charged particle multiplicity which is
dominated by protons, $\pi^{\pm}$ and $K^{\pm}$ (cf. Fig. 5).
However, above about 50 -- 100 A$\cdot$GeV the meson ratios do not
change very much (cf. Figs. 5,6) while the charged particle
multiplicity becomes dominated by $\pi^{\pm}$ and $K^{\pm}$. Thus from 
SPS to RHIC energies the low mass dilepton yield should
approximately be proportional to the charged particle
multiplicity, too.

The relative change in the dilepton spectra is quantitatively displayed
in Fig. \ref{Fig9} -- again for central ($b$= 2 fm) collisions of
Au~+~Au -- for the case of free meson spectral functions (solid lines) and
the spectral function from Rapp et al.  \cite{RappNPA,CasRap} (dashed
lines). In line with the discussion above the most prominent spectral
changes are expected at rather low bombarding energies from
2--10~A$\cdot$GeV, where the enhancement in the invariant mass range from 
0.3--0.6~GeV is about a factor of 4. Note that the $\omega$ and
$\phi$ vacuum decays show clear peaks on top of the 'exponential'
continuum, which will have to be identified experimentally. We mention
that in all our calculations we have incorporated an experimental
('optimistic') mass resolution of $\Delta M$ = 10 MeV for the dilepton
invariant mass.

\section{Charmonium production and suppression}

Matsui and Satz \cite{matsui} have proposed that a suppression of the
$J/\Psi$ yield in ultra-relativistic heavy-ion collisions is a
plausible signature for the formation of the quark-gluon plasma because
the $J/\Psi$ should dissolve in the QGP due to color screening.  This
suggestion has stimulated a number of heavy-ion experiments at CERN SPS
to measure the $J/\Psi$ via its dimuon decay. Indeed, these experiments
have shown a significant reduction of the $J/\Psi$ yield when going from 
proton-nucleus to nucleus-nucleus collisions \cite{NA38}.
Especially for Pb~+~Pb at 160~A$\cdot$GeV an even more dramatic
reduction of $J/\Psi$ has been reported by the NA50 Collaboration
\cite{NA50,gonin,NA5099}.

To interpret the experimental results, various models based on
$J/\Psi$ absorption by hadrons have been also proposed (cf. Refs.
\cite{Cass99,Vogt99,Seattle98} for recent reviews) that do not
involve the assumption of a QGP phase transition. The role of
comover dissociation is presently again heavily debated
\cite{Bernd,Seattle98} especially since theoretical calculations
for $J/\Psi$-meson dissociation cross sections differ by up to a
factor of 50 \cite{Bernd,Haglin,Bernd96,Blaschke,DimaSatz}. The
problem is even more complicated since the $J/\Psi$ meson is not
created instantly as a hadronic state and there is also a
substantial ($\approx$ 35 \%)  feeding from the $\chi_c -\gamma$
decays. Moreover, it is expected that the $c\bar{c}$ pair is first
produced in a color-octet state together with a gluon
('pre-resonance state') and that this more extended configuration
has a larger interaction cross section with baryons and mesons
before the $J/\Psi$ singlet state, $\Psi'$ or $\chi_c$ finally
emerges after some formation time $\tau_c$. Additionally, the
dissociation on mesons of formed $J/\Psi$'s will differ from
$\chi_c$ due to their different thresholds with respect to the
$D\bar{D}$ channel as well as for the $\Psi^{\prime}$. Since
experimental information on the various charmonium-meson cross
sections -- especially at low relative momenta -- will be hard to
obtain, the excitation function of charmonium suppression in
central nucleus-nucleus collisions might be exploited to obtain
additional information on the absorption scenarios. One expects
that quite below the bombarding energy necessary for the formation
of a QGP phase the charmonium absorption should be entirely due to
dissociation with hadrons; any additional suppression due to color
screening then will show up in a more rapid suppression with the
incident energy or with the energy density achieved
\cite{SatzQM99}.

In this Section we calculate the excitation function for $J/\Psi$
suppression within two absorption scenarios, i.e. the 'early' and
'late' comover dissociation models, which have been explored in detail
by our group before at SPS energies \cite{CaKo97JPsi,Cass97d,Geiss2}.
The 'early' comover absorption scenario is based on the idea, that a $c
\bar{c}$ pair is created in the initial 'hard' phase of the
nucleus-nucleus collision, where the string density is very high, and
the $c\bar{c}$ pair is dissolved in the color electric field of
neighboring strings.  Since in the HSD approach the information on the
string density as well as the string space-time extension is available,
the absorption model has only a single parameter, i.e. the average
transverse dimension of an extending string. In Ref. \cite{Geiss2} a
string radius of 0.2 fm was found to describe simultaneously the data
for p~+~A and S~+~U at 200 A$\cdot$GeV from NA38 \cite{NA38} and for
Pb~+~Pb at 160 A$\cdot$GeV from NA50 \cite{NA50} when adopting the
conventional charmonium-nucleon dissociation cross section of 6 mb. It
has been also speculated that the overlap of strings due to percolation
might describe the phase transition to a QGP \cite{Satz99}.

In the 'late' comover scenario the additional charmonium suppression is
due to charmo\-nium-meson scattering to $D\bar{D}$ with an average
charmonium-meson cross section of $\approx$ 3 mb \cite{Cass97d}. Cross
sections of this order have been calculated by Haglin in Ref.
\cite{Haglin} within a meson-exchange model and thus might appear not
unrealistic. In our present calculation we refer to the model II of
Ref. \cite{Cass97d} including a 'pre-resonance' charmonium life time of
0.3--0.5~fm/c which is supported (within the errorbars) by a more
recent analysis of charmonium suppression as a function of the Feynman
variable $x_F$ in p~+~A reactions \cite{pAdata} from He et al.
\cite{Huefner} and Kharzeev et al. \cite{Dima99}. We recall that within
the model II of Ref. \cite{Cass97d} the $J/\Psi$ suppression data for
p~+~A and S~+~U at 200 A$\cdot$GeV from NA38 \cite{NA38} and for Pb~+~Pb
at 160 A$\cdot$GeV from NA50 \cite{NA50} have been described very well
when adopting a 'pre-resonance'-nucleon cross section of 6 mb and a
$J/\Psi$-nucleon cross section of 3--4~mb in line with the data on
$J/\Psi$ photoproduction \cite{Photo}. Independent dynamical studies on
charmonium suppression within the UrQMD model \cite{Gerland,Spieles}
later on have lead to very similar conclusions.

Since the comparison of our calculations at SPS energies has been
performed to data taken in 1995 and before we show in Fig.
\ref{Fig10} a comparison of the 'early' comover model (dashed
line) \cite{Geiss2} and the 'late' comover model II \cite{Cass97d}
with the more recent data from NA50 \cite{NA5099,NA50QM} for
Pb~+~Pb at 160 A$\cdot$GeV using the explicit numbers from Refs.
\cite{Cass97d,Geiss2}\footnote{This comparison is necessary since
the 1995 data have been rescaled in \cite{NA50QM} and our
calculations are reported inconsistently in the more recent
presentations of this topic \cite{NA50QM,BM99}.}. The 'early'
comover absorption model here gives a little too low suppression
at high $E_T$ whereas the 'late' comover absorption model is still
in line with the more recent data from 1996 and 1998 with minimum
bias (open triangles and open circles). The data from 1996 (full
squares), that show a (much debated) two-step behaviour, do not
agree with the explicit $E_T$ dependence from our calculations;
however, the 1996 minimum bias data (open triangles) well match
for $E_T \geq$ 40 GeV whereas the $J/\Psi$ suppression is slightly
overestimated at lower $E_T$. We do not comment on the highest
$E_T$ data points from 1998 with minimum bias since our earlier
analysis did not extend to these specific events.

The question now arises, if the excitation function for $J/\Psi$
suppression in central Au~+~Au collisions might show some unusual
behaviour within the two scenarios discussed above or how they
might be disentangled. In order to achieve the same suppression
factor in central collisions within the 'early' comover model we
have increased the string absorption radius from $r_s$ = 0.2 fm to
0.22 fm to get the same value for the $J/\Psi$ survival factor
$S_{J/\Psi}$ at 160 A$\cdot$GeV. We note that the total $J/\Psi$
multiplicity shown in Fig.  \ref{Fig5} has been calculated within
the 'late' comover model. It drops by almost 3 orders of magnitude
when decreasing the bombarding energy from 160 A$\cdot$GeV to 20
A$\cdot$GeV.  At the lowest energy considered here the
experimental $J/\Psi$ signal will be very hard to measure; it is
hopeless within the present experimental setups. Nevertheless, it
is worth exploring theoretically if some unusual excitation
function might be found.  Our results are displayed in Fig.
\ref{Fig11} (l.h.s.) for the 'early' absorption model (open
circles) and for the 'late' comover model (full circles); both
models practically do not differ in their excitation functions and
show a very smooth decrease of $S_{J/\Psi}$ from 0.4 to 0.3 with
increasing bombarding energy.  The net absorption by baryons is
dominant in both scenarios, however, differs in magnitude due to
the simple fact that in the 'early' (string) absorption model
there is less suppression by baryons since the absorption by
strings competes at early times.  In the 'late' comover model
there is more absorption by baryons because the mesons are formed
at later stages and not competitive in the early phase; their
relative contribution is lower as for strings accordingly.
However, these individual contributions cannot be distinguished
experimentally and thus are 'irrelevant'.

The r.h.s. of Fig. 11 shows the survival probability $S_{J/\Psi}$
in the 'late' comover model for central Au + Au collisions from
0.160 A$\cdot$TeV to 21.5 A$\cdot$TeV, respectively, where we have gated on
$J/\Psi$'s in the rapidity interval -1 $\leq y_{cm} \leq$ 1. The
solid line stands for the total $J/\Psi$ survival probability
while the dashed line displays the relative absorption on baryons
and the dotted line the relative dissociation on mesons. Whereas
the dissociation on baryons is practically constant with
bombarding energy, the absorption on mesons increases in line with
the higher meson densities achieved with increasing $E/A$. We note
that the 'early' comover model leads to a similar total absorption
within the numerical accuracy.

We have to mention that neither the 'early' nor the 'late' comover
model might be realized in nature exclusively and both absorption
processes should occur within the same reaction with probably
different weights. Since we do not find a substantial difference
for both scenarios also a linear combination of both absorption
models, i.e.  decreasing $r_s$ as well as the charmonium-meson
cross section accordingly, will lead to a similar excitation
function. This also holds for the relative suppression on the
transverse energy $E_T$ (cf. Fig. \ref{Fig10}).

Inspite of the rather disappointing perspectives to disentangle
the 'late' and 'early' comover models experimentally at the full
range of SPS energies, the excitation function of $J/\Psi$ still
might show some discontinuity in $E/A$ experimentally, which could
rule out the two models studied here and indicate a transition to
a QGP phase. This subject is taken up in the next Section again
with respect to the dependence of $S_{J/\Psi}$ on the transverse
energy $E_T$ produced in Au + Au collisions at RHIC energies.

\section{Predictions for RHIC energies}

As noted in the introduction one expects that the initial
nonequilibrium phase of a nucleus-nucleus collision at RHIC
energies should be described by parton degrees of freedom, whereas
hadrons are only formed (by 'condensation') at a later stage of
the reaction and interact until freeze out.  Thus parton cascade
calculations should be adequate for all initial reactions
involving a large 4-momentum transfer between the constituents,
while hadron cascades should be appropriate in the final hadronic
expansion phase. We suggest that the dynamics in between the
partonic and hadronic phase might be described by quarks
(diquarks) and strings as e.g. implemented in the HSD approach.

The practical question is, however, if nonequilibrium partonic and
hadron/string models can be distinguished at all, i.e. do they
lead to different predictions for experimental observables?  In
fact, first applications of the parton cascade model developed by
Geiger \cite{Geiger1,VNI} to nucleus-nucleus collisions at SPS
energies \cite{Geiger} have suggested that a reasonable
description of the meson and baryon rapidity distributions can
also be achieved on the basis of partonic degrees of freedom.
However, in the latter calculations the extrapolation of the
strong coupling constant to low $Q^2$ has been overestimated as
discovered recently by Bass and M\"uller \cite{Bass99c}. This
finding invalidates the detailed predictions and comparisons
within the hybrid models VNI+UrQMD or VNI+HSD presented in Ref.
\cite{QMRHIC} that have been tailored to describe the dynamics at
RHIC or even LHC energies.

We start with $pp$ collisions at $\sqrt{s}$ = 200 GeV.  The
calculated results for the proton, $\pi^+$ and $K^+$ rapidity
distributions in the cms are shown in Fig. \ref{Fig12} (upper
part) for both models, which are denoted individually by the
labels VNI and HSD in obvious notation. On the level of $pp$
collisions we find only minor differences between the two kinetic
models. The parton cascade shows a slightly higher amount of
proton stopping as the HSD model (l.h.s.) and as a consequence a
slightly higher production of $\pi^+$ and $K^+$ mesons (r.h.s.),
because the energy taken from the relative motion of the leading
baryons is converted to the production of mesons. It is presently
unclear which of the two approaches will be closer to experiment;
a proper description of $pp$ data will be a necessary step before
performing reliable extrapolations to nucleus-nucleus collisions.

Inspite of this missing experimental information we directly step
towards central collisions ($b \leq$ 1.5 fm) for Au~+~Au at
$\sqrt{s}$ = 200~GeV. The calculated results for the net proton
(here $p-\bar{p}$), antiproton, $\pi^+$ and $K^+$ rapidity
distributions in the cms are shown in Fig. \ref{Fig12} (lower
part). In the HSD scenario essentially 'comover' scattering occurs
with a low change of the meson rapidity distribution. Thus the
meson rapidity distributions are roughly the same as for $pp$
collisions. Also note that at midrapidity the net baryon density
$\sim N_p - N_{\bar{p}}$ is practically zero, however, even at
midrapidity at lot of baryons appear that are produced together
with antibaryons. Thus also mesons (especially $c \bar{c}$ pairs)
encounter a lot of baryons and antibaryons on their way to the
continuum. Whereas the HSD approach predicts a vanishing net
baryon density at midrapidity, other recent models -- that combine
high and low energy transport concepts -- suggest a sizeable net
proton density for $y_{cm} \approx$ 0 \cite{Ko99}.

The amount of higher order hadronic rescattering processes at RHIC
energies is depicted in Fig. \ref{Fig13} (lower right part) as
emerging from the HSD calculation, where the number of
baryon-baryon ($BB$) and meson-baryon collisions ($mB$) is shown
as a function of the invariant energy $\sqrt{s}$.  We mention that
quark-baryon and diquark-baryon collisions are counted here as
$mB$ or $BB$ collisions, respectively. Apart from the initial
small peak at $\sqrt{s}$ = 200 GeV a substantial amount of
intermediate and low energy rescattering processes with maxima at
2.5 GeV and 1.8 GeV are found, which essentially stand for flavor
exchange processes, multiple pion production in $mB$ and $BB$
collisions as well as secondary strangeness production channels.
For comparison the corresponding $\sqrt{s}$ distributions are also
displayed for bombarding energies of 2, 11 and 160 A$\cdot$GeV,
respectively, showing a dominance of low energy $BB$ and $mB$
collisions. The latter reactions occur at energy scales where
perturbative QCD is no longer applicable. This has to be kept in
mind additionally when comparing to $pp$ and $pA$ reactions at
$\sqrt{s}$ = 200 GeV.

We return to the question of charmonium suppression at RHIC
energies since it is expected that one might probe increasing
energy densities also with increasing centrality of the collision,
where the latter can be correlated with the transverse energy
$E_T$ produced in a collision event. As argued e.g. by Satz
\cite{SatzQM99} the survival factor $S_{J/\Psi}$ then should show
steps as a function of $E_T$ due to the melting of first the
$\chi_c$ and then the $J/\Psi$ in a QGP phase. As seen from Fig.
10 there are no pronounced steps in the $E_T$ dependence of
$J/\Psi$ suppression in the data for Pb + Pb at SPS energies
according to the authors point of view; this situation might
change at RHIC energies.

Using the 'late' comover model described in Section 5 we have
calculated the $J/\Psi$ survival factor $S_{J/\Psi}$ as a function
of the transverse energy $E_T$ in the cms rapidity window [-1,1]
for Au + Au at $\sqrt{s}$ = 200 GeV on an event by event basis
covering all impact parameters from $b$ = 0 to 13 fm. The resulting
correlation of $S_{J/\Psi}$ with $E_T$ is shown in Fig. 14 and
indicates a smooth decrease with centrality (or increasing $E_T$)
reaching an average survival probability of $\approx$ 0.1 for the
most central events (cf. Fig. 11, r.h.s.). This result can be
understood as follows: According to our calculations the net
$J/\Psi$ dissociation by mesons in central Au + Au collisions at
the SPS is $\approx 16\%$ (cf. Fig. 11) while the rapidity
distribution of negatively charged particles ($h^-$) at
midrapidity here is about 180. At the RHIC energy we get a
corresponding $h^-$ rapidity density at midrapidity of $\approx$
450 (cf. Fig. 12) which is higher by a factor of 2.5. Simply
multiplying the $J/\Psi$ meson absorption at the SPS of $16 \%$ by
the factor 2.5 we obtain about $40 \%$ for central collisions at
RHIC energies, which together with $\approx 52 \%$ of absorption
on baryons gives a survival probability of $8 \%$. The actual
numerical results in Fig. 14 indicate that this simple estimate
works quite well. On the other hand, if the $h^-$ rapidity density
is found to be lower (higher) experimentally, we expect
corresponding changes in the $J/\Psi$ suppression for central
events if the 'late' comover absorption model holds true. This
dependence might well be tested experimentally in the near future
to possibly falsify the comover dissociation model.

\section{Summary}

\begin{figure}[t]
\phantom{a}\vspace*{-3cm}
\centerline{\psfig{figure=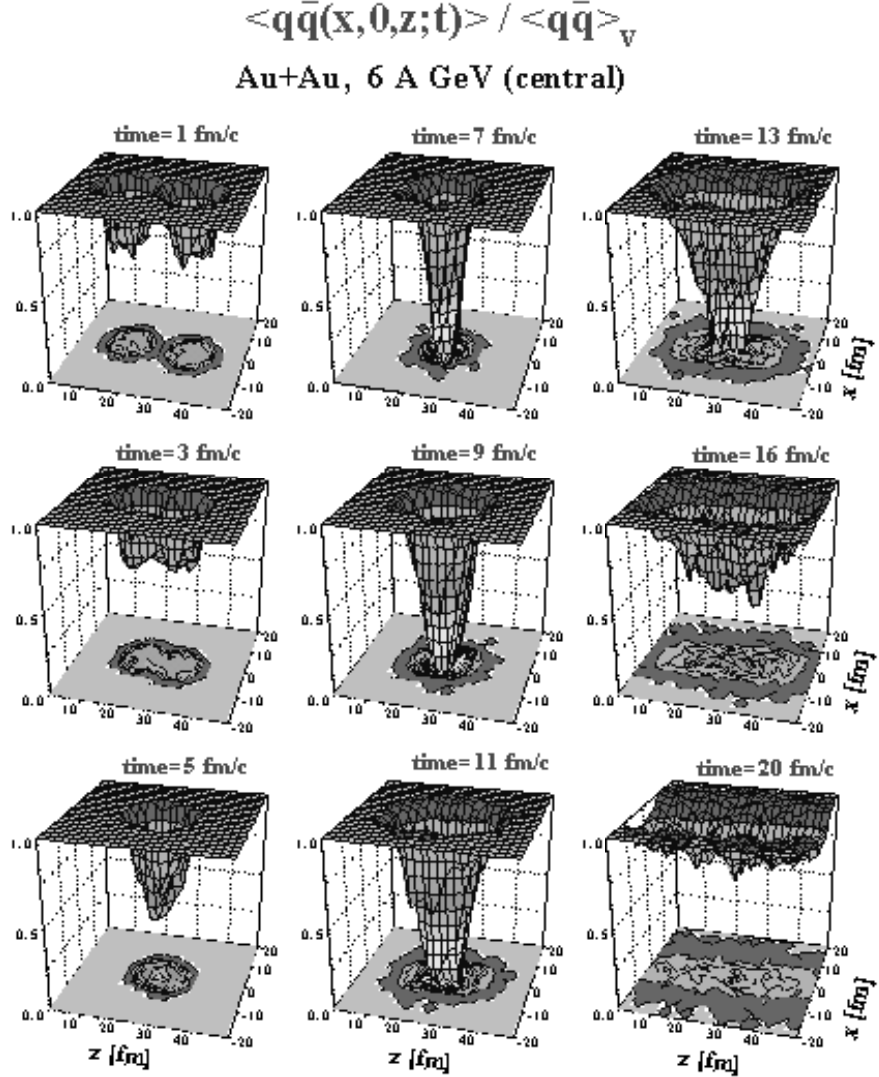,width=16cm}} \vspace*{-2cm}
\caption{The scalar quark condensate $<q\bar{q}(x,0,z;t)>$ for
central Au~+~Au collisions at 6 A$\cdot$GeV divided by the vacuum
condensate $<q\bar{q}>_V$ such that the nonperturbative vacuum is
characterized by a value of 1. The $z$-direction has been
stretched by the Lorentz-factor $\gamma_{cm}$ to compensate for
Lorentz contraction, while negative numerical values for the
condensate have been suppressed.} \label{Fig3}
\end{figure}

\begin{figure}[t]
\phantom{a}\vspace*{-2.5cm}
\centerline{\psfig{figure=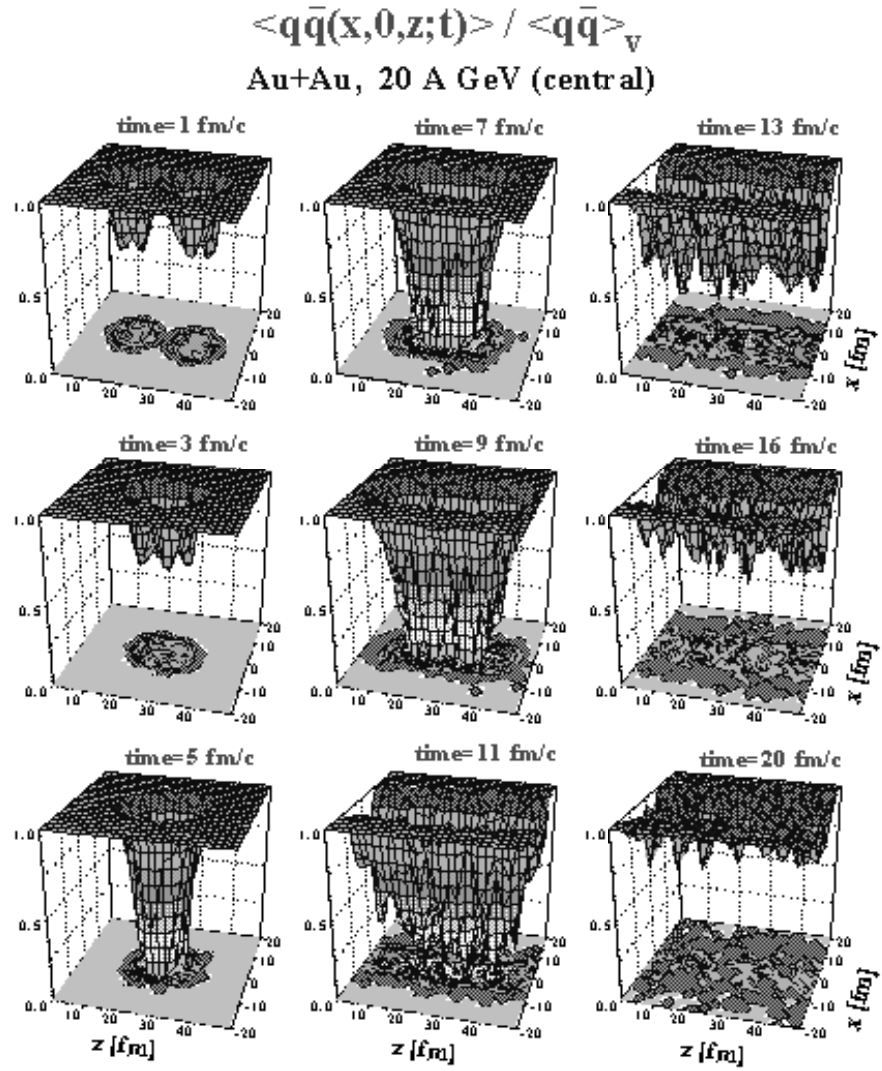,width=16cm}}
\vspace*{-2cm}
\caption{Same as Fig. 3 for central collisions of Au~+~Au at 20 A$\cdot$GeV.}
\label{Fig4}
\end{figure}

In this work we have performed a systematic analysis of hadron
production in central Au~+~Au collisions from SIS to RHIC energies
within the HSD transport approach. We have concentrated here on the
'classical' signatures, i.e. strangeness and low mass dilepton
enhancement as well as charmonium suppression. For all observables our
calculations give a monotonic increase (for the ratio $K^+/\pi^+$ and
charmonium suppression) or decrease (for the low mass dilepton
enhancement) with bombarding energy, respectively.  So far, experimental
data are available only in a limited range of bombarding energies or at
a single energy, respectively. We have pointed out that the relative
maximum indicated by the experimental data in the $K^+/\pi^+$ ratio at
about 10 A$\cdot$GeV (or higher?) is not reproduced within the transport
approach that is based on quark, diquark, string and hadronic degrees
of freedom.  We speculate that at AGS energies this failure might be
attributed to a restoration of chiral symmetry in a sufficiently large
space-time volume (cf. Figs. \ref{Fig3} and \ref{Fig4}).

\begin{figure}[t]
\phantom{a}\vspace*{-2.5cm}
\centerline{\psfig{figure=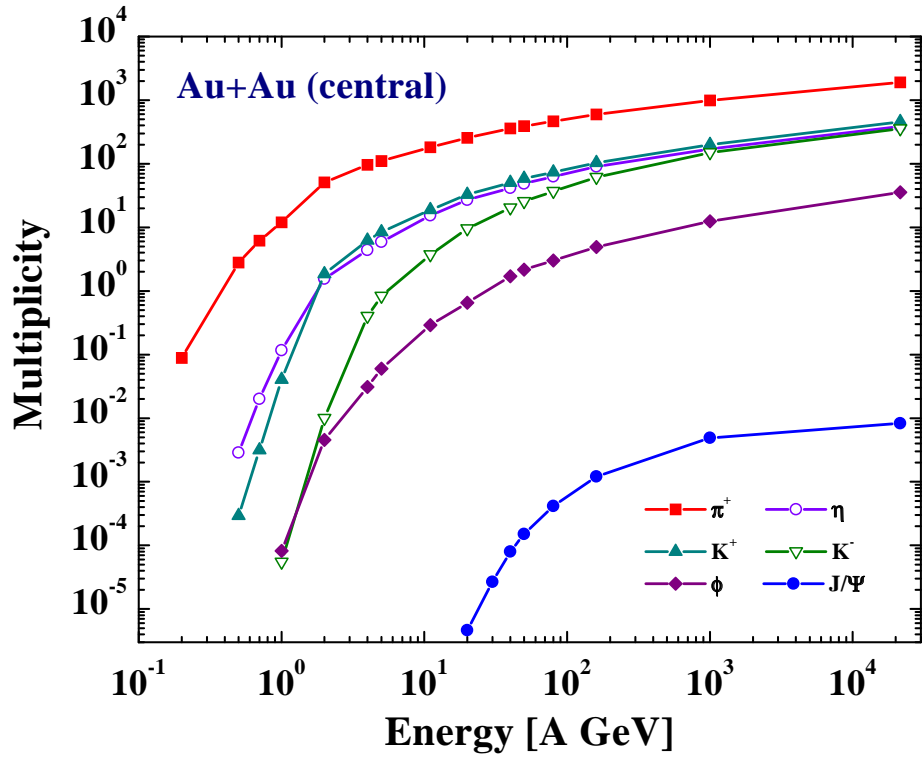,width=16cm}}
\vspace*{-5cm}
\caption{The meson ($\pi^+, \eta, K^+, K^-, \phi$ and $J/\Psi$)
multiplicities from the HSD approach for central collisions of $Au
+ Au$ from 200 A MeV to 21.5 A$\cdot$TeV.}
\label{Fig5}
\end{figure}

\begin{figure}[t]
\phantom{a}\vspace*{-2.5cm}
\centerline{\psfig{figure=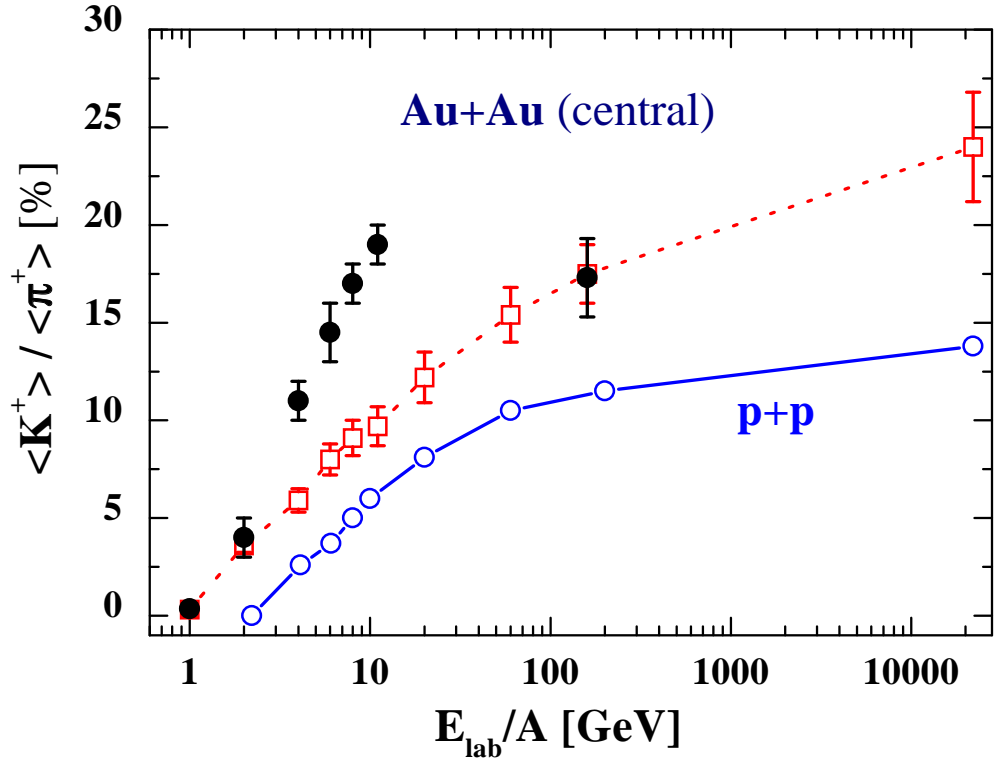,width=16cm}}
\vspace*{-5cm}
\caption{The $K^+/\pi^+$ ratio at midrapidity from central Au~+~Au ($Pb
+ Pb$) collisions from 1 A$\cdot$GeV to 21.5 A$\cdot$TeV. The open circles
show the results from HSD for $pp$ collisions while the open
squares are obtained for central Au~+~Au reactions. The
experimental data from Refs. \protect\cite{QM99,E866,Alard} are
displayed in terms of the full circles.}
\label{Fig6}
\end{figure}

\begin{figure}[t]
\phantom{a}\vspace*{-2.5cm}
\centerline{\psfig{figure=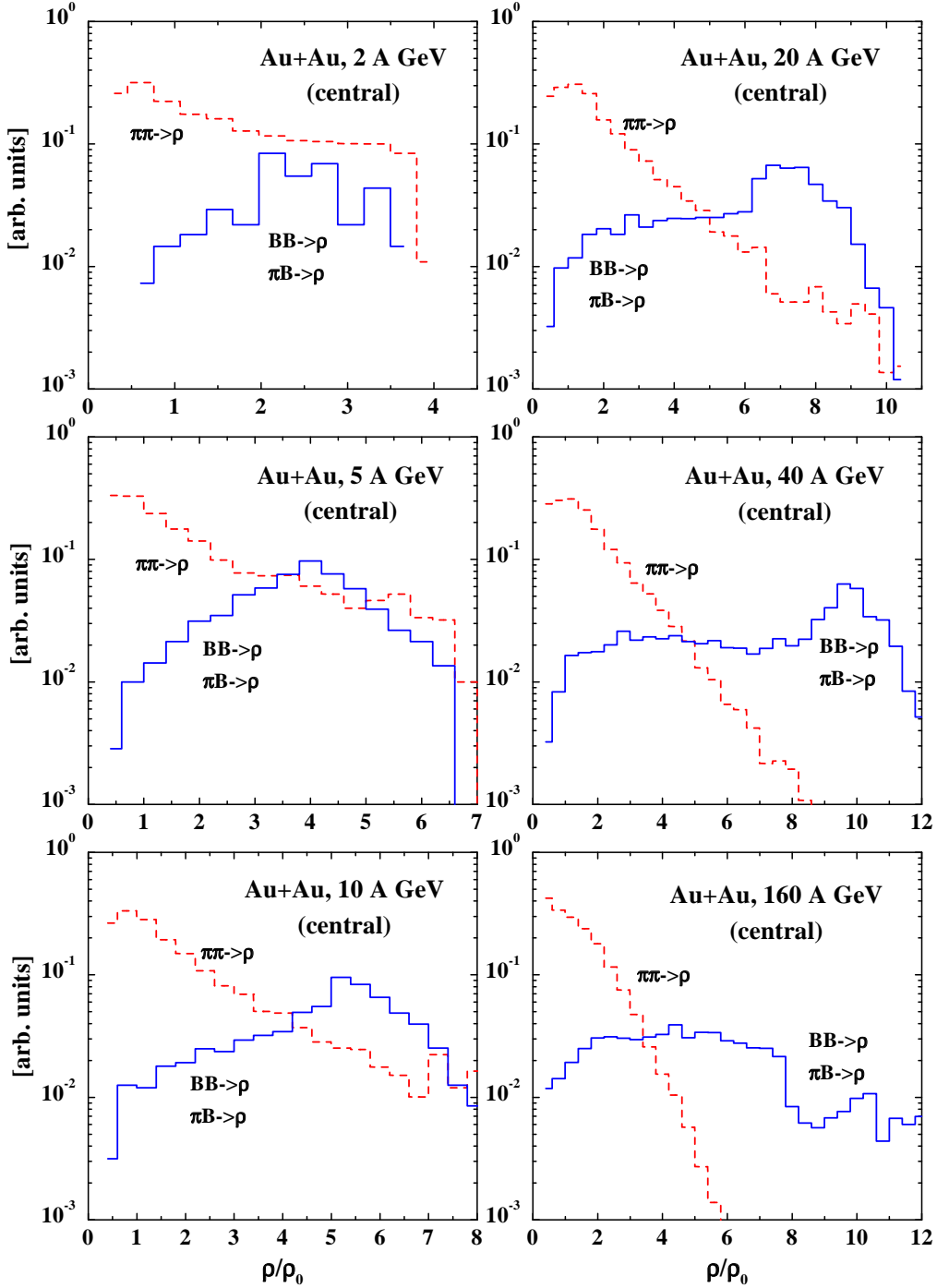,width=16cm}}
\vspace*{-2cm}
\caption{The differential $\rho$-meson distribution versus baryon density
$\rho/\rho_0$ (at the $\rho$ creation point) for central
collisions of Au~+~Au at 2, 5, 10, 20, 40, and 160 A$\cdot$GeV from
the HSD approach. The production channels involving two baryons or
a meson and a baryon (denoted by $BB \to \rho, \pi B \to \rho$)
are summed up in the solid histograms whereas the dashed
histograms stand for the sum of the meson production channels
(denoted by $\pi\pi \to \rho$) which are dominated by the pion
annihilation channel.}
\label{Fig7}
\end{figure}

\begin{figure}[t]
\phantom{a}\vspace*{-2.5cm}
\centerline{\psfig{figure=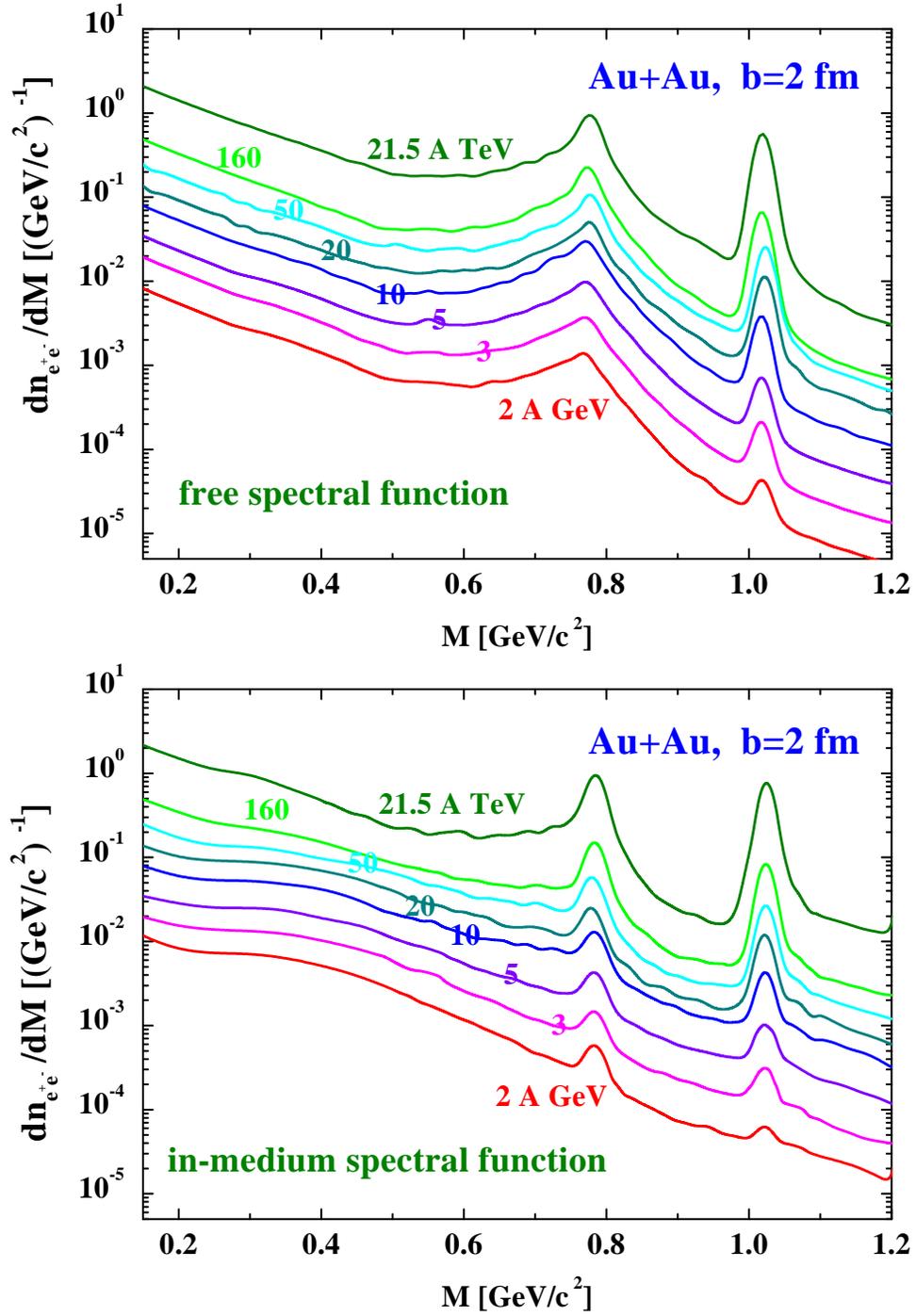,width=16cm}} \vspace*{-2cm}
\caption{The differential dilepton multiplicity $d n_{e^+e^-}/dM$
for central collisions of Au~+~Au for 2, 3, 5, 10, 20, 50, 160,
and 21500 A$\cdot$GeV. Upper part: HSD calculations involving the
'free' $\rho$-meson spectral function: lower part: HSD calculation
involving the in-medium $\rho$ spectral function from Rapp et al.
\protect\cite{RappNPA,CasRap}.} \label{Fig8}
\end{figure}

\begin{figure}[t]
\phantom{a}\vspace*{-2.5cm}
\centerline{\psfig{figure=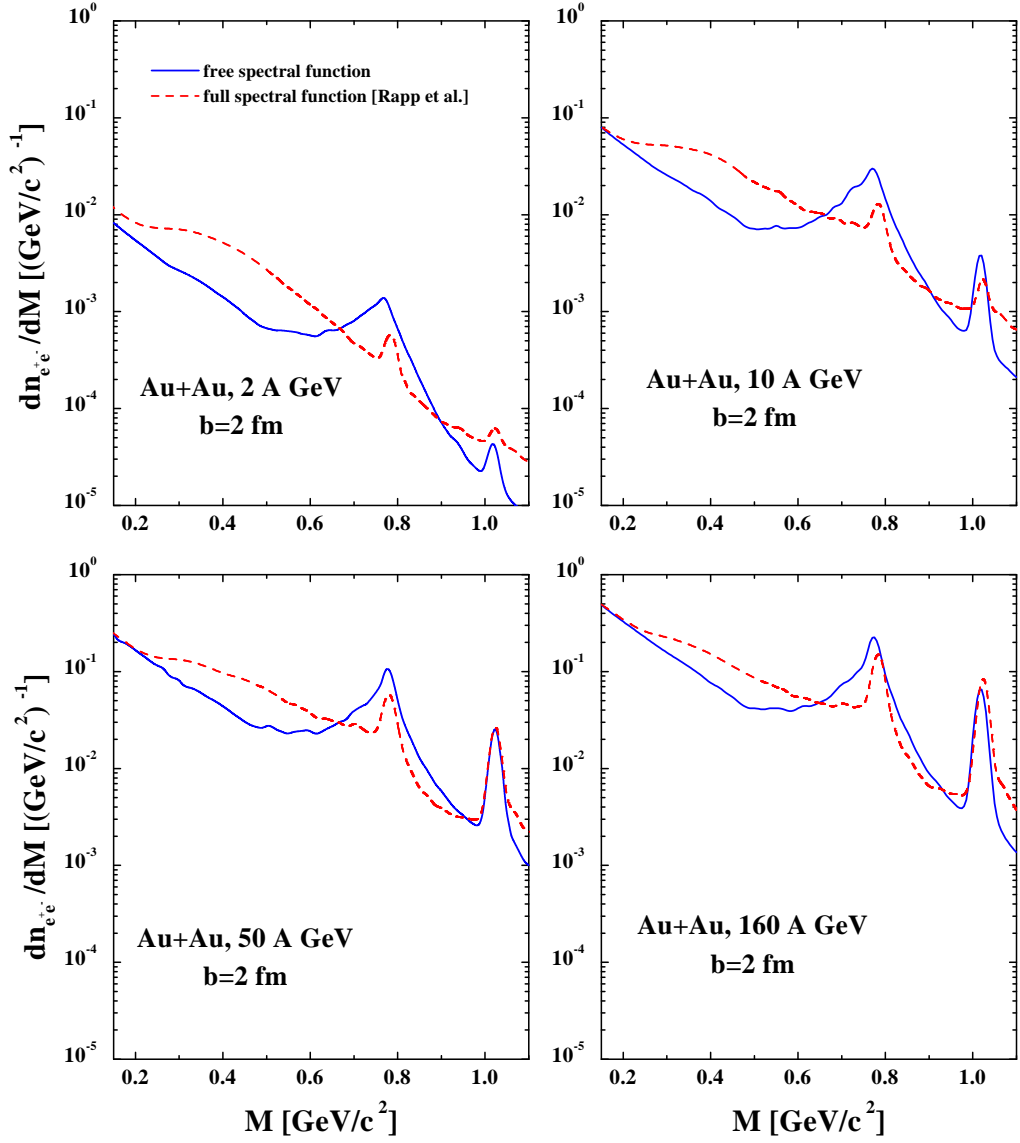,width=16cm}} \vspace*{-3cm}
\caption{The differential dilepton multiplicity $d n_{e^+e^-}/dM$
for central collisions of Au~+~Au at 2, 10, 50, and 160
A$\cdot$GeV from the HSD calculations involving the 'free'
$\rho$-meson spectral function (solid lines) and the in-medium
$\rho$ spectral function from Rapp et al.
\protect\cite{RappNPA,CasRap} (dashed lines) for comparison.}
\label{Fig9}
\end{figure}

\begin{figure}[t]
\phantom{a}\vspace*{-2.5cm}
\centerline{\psfig{figure=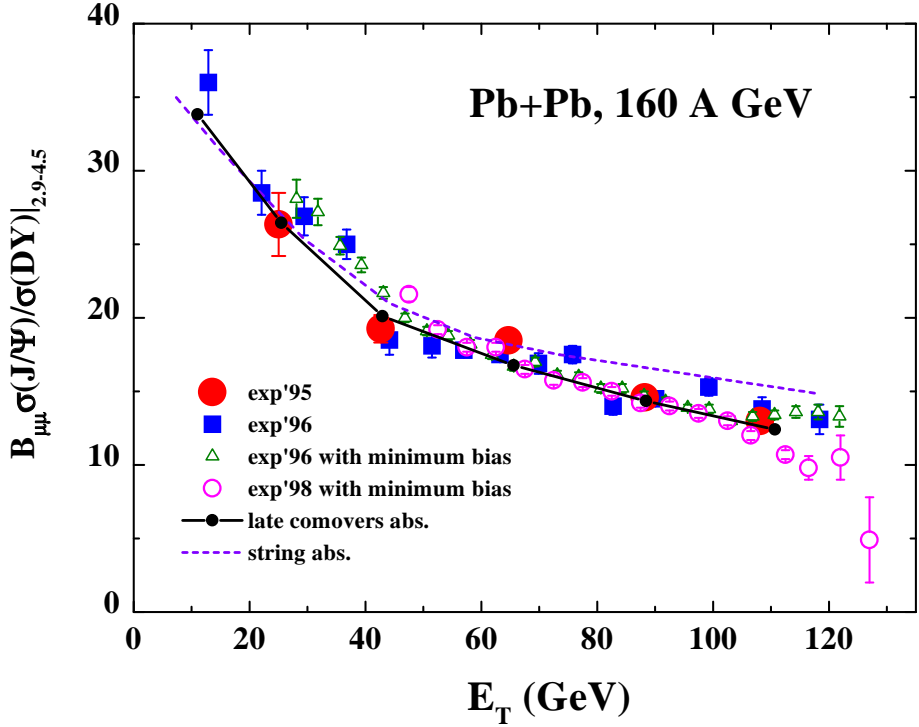,width=16cm}}
\vspace*{-5cm}
\caption{The $J/\Psi$ suppression (in terms of the $\mu^+ \mu^-$ decay
branch relative to the Drell-Yan background from 2.9 -- 4.5 GeV
invariant mass) as a function of the transverse energy $E_T$ in
Pb~+~Pb collisions at 160 A$\cdot$GeV. The solid line stands for the
HSD result within the 'late' comover absorption scenario from Ref.
\protect\cite{Cass97d} while the dashed line results from the
'early' string absorption scenario from Ref. \protect\cite{Geiss2}
involving a transverse string radius $r_s$ = 0.2 fm. The full dots
stand for the NA50 data from 1995 \protect\cite{NA50}, the full
squares for the 1996 data \protect\cite{NA5099}, the open
triangles for the 1996 data with minimum bias
\protect\cite{NA5099} while the open circles represent the 1998
data \protect\cite{NA50QM}. Note that the 1995 data have been
rescaled in $E_T$ as compared to Ref. \protect\cite{NA50}; the
same rescaling has been adopted to the calculations from Refs.
\protect\cite{Cass97d,Geiss2} which had been compared to the 
data from \protect\cite{NA50}.}
\label{Fig10}
\end{figure}

\begin{figure}[t]
\phantom{a}\vspace*{-2.5cm}
\centerline{\psfig{figure=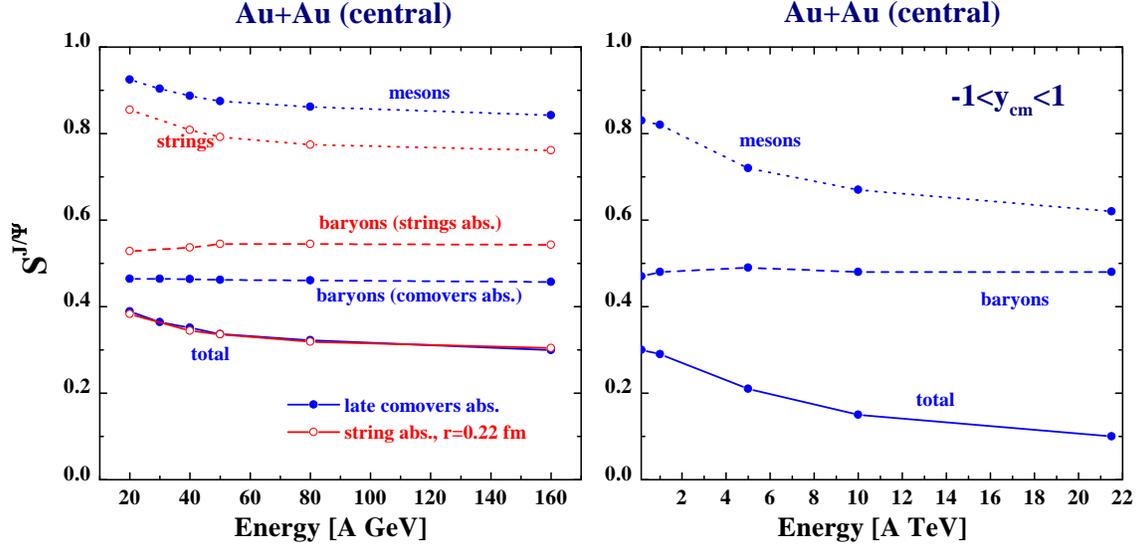,width=16cm}} \vspace*{-5cm}
\caption{The $J/\Psi$ survival factor $S^{J/\Psi}$ (in terms of
the $\mu^+ \mu^-$ decay branch relative to the Drell-Yan
background from 2.9 -- 4.5 GeV invariant mass normalized to the
same quantity for $pd$ reactions) as a function of the bombarding
energy in central Au~+~Au collisions from 20 to 160 A$\cdot$GeV
(l.h.s.) and from 160 A$\cdot$GeV to 21.5 A$\cdot$TeV (r.h.s.).
The solid line (full dots) stands for the HSD result within the
'late' comover absorption scenario from Ref.
\protect\cite{Cass97d} while the solid line (open circles) results from 
the 'early' string absorption scenario from Ref.
\protect\cite{Geiss2} involving a transverse string radius $r_s$ =
0.22 fm in order to match both absorption scenarios at 160
A$\cdot$GeV. The dashed lines stand for the relative fraction of
$J/\Psi$ dissociations with baryons while the dotted lines stand
for the 'early' comover string dissociation (open circles) and
'late' comover meson dissociation (full dots), respectively. The
total suppression factors (full lines) are practically identical
for both scenarios from 20 to 160 A$\cdot$GeV (l.h.s.). This also
holds for the energy range from 160 A$\cdot$GeV to 21.5
A$\cdot$TeV within the numerical accuracy achieved. }
\label{Fig11}
\end{figure}

\begin{figure}[t]
\phantom{a}\vspace*{-2.5cm}
\centerline{\psfig{figure=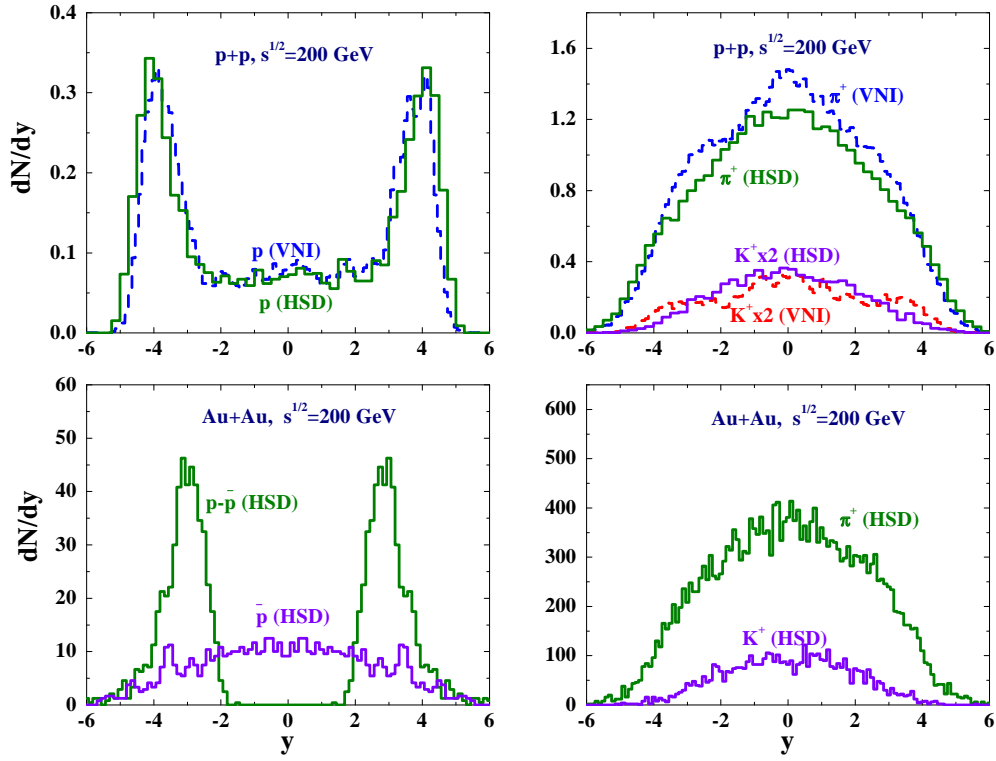,width=16cm}} \vspace*{-5cm}
\caption{The rapidity distribution of protons (upper left part) from 
$pp$ collisions at $\sqrt{s}$ = 200 GeV from the HSD approach
(solid histogram) in comparison to the prediction from the parton
cascade VNI \protect\cite{VNI} (dashed histogram); the upper right
part shows the same comparison for the $\pi^+$ and $K^+$ rapidity
distributions, respectively. The lower part of the figure displays
the HSD predictions for central ($b \leq$ 1.5 fm) Au~+~Au at
$\sqrt{s}$ = 200 GeV per nucleon; (l.h.s.) net proton
($p-\bar{p}$) and $\bar{p}$ rapidity distribution, (r.h.s.)
$\pi^+$ and $K^+$ rapidity distributions.} \label{Fig12}
\end{figure}

\begin{figure}[t]
\phantom{a}\vspace*{-2.5cm}
\centerline{\psfig{figure=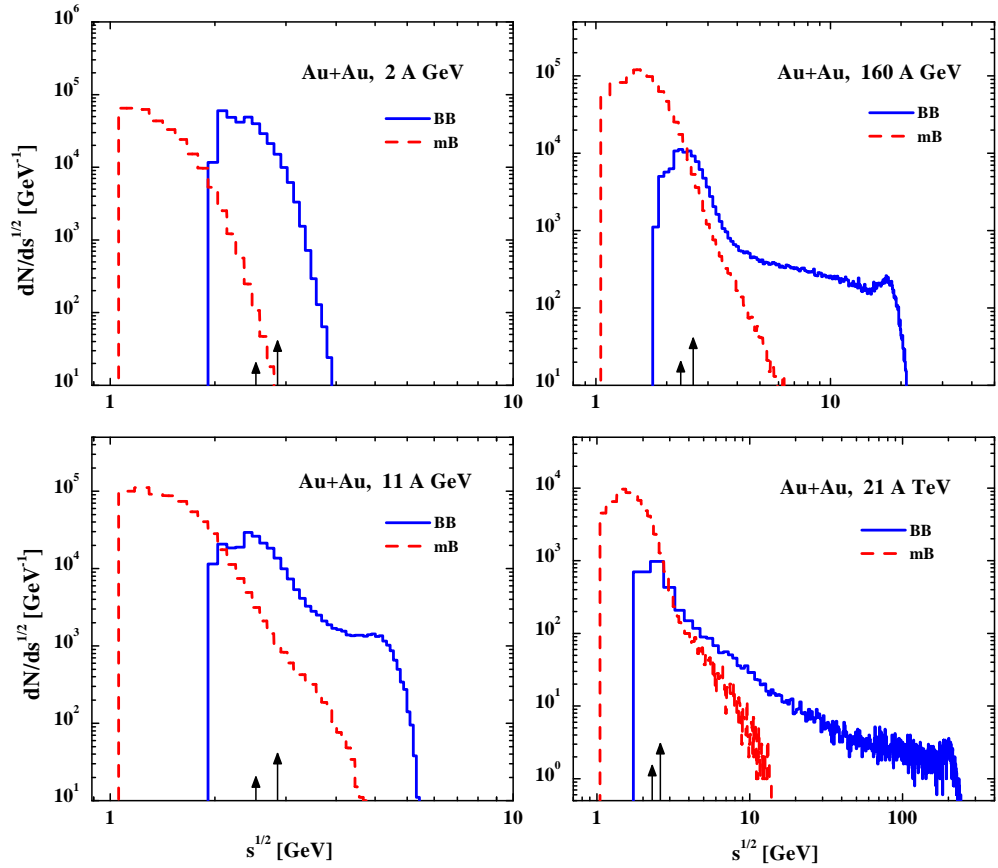,width=16cm}} \vspace*{-5cm}
\caption{The number of baryon-baryon (solid histograms) and
meson-baryon collisions (dashed histograms) as a function of the
invariant energy $\sqrt{s}$ for central ($b \leq$ 1.5 fm) Au~+~Au
collisions at bombarding energies of 2, 11, 160 A$\cdot$GeV and 21
A$\cdot$TeV, respectively, from the HSD model. The arrows indicate the
string thresholds for $mB$ and $BB$ collisions of 2.3 GeV and 2.6
GeV, respectively.} \label{Fig13}
\end{figure}

The enhancement of low mass dileptons in the range 0.3 GeV $\leq
M_{e^+e^-} \leq$  0.6 GeV is most pronounced at lower bombarding
energies of 2--5 A$\cdot$GeV within our calculations since here the
space-time volume for densities above 2 $\rho_0$ is very large such
that a majority of $\rho$-mesons probes the high density phase of the
reaction (cf. Fig. \ref{Fig5}). With increasing bombarding energy the
average density -- which a $\rho$-meson experiences -- drops
substantially such that high energy nucleus-nucleus collisions are not
well suited for in-medium $\rho$ spectroscopy.

The suppression of charmonium (here $J/\Psi$) increases smoothly
with bombarding energy and centrality of the reaction within the
'early' and 'late' comover absorption scenarios. Unfortunately,
both scenarios cannot be distinguished by means of the excitation
function since they give approximately the same survival
probability $S_{J/\Psi}$ with bombarding energy (cf. Fig. 11). In
the transport approach the smooth increase of charmonium
absorption with bombarding energy is easy to understand:  a major
fraction of $J/\Psi$'s is anyhow dissociated by baryons which
basically are of the same number at all energies considered here;
only the relative collisional energy changes. The additional
absorption by 'early' strings or 'late' hadrons increases smoothly
with bombarding energy since the string and hadron density
increases accordingly. At RHIC energies this additional
suppression mechanism leads to a $J/\Psi$ suppression of about
90\% in central Au~+~Au collisions even without employing an
explicit formation of a QGP. We note, however, that the charmonium
suppression shows a smooth dependence on the transverse energy
$E_T$ (cf. Fig. 14); any gradual steps of $S_{J/\Psi}$ with $E_T$
due to a melting of the $\chi_c$ or the $J/\Psi$ at higher energy
density would indicate a new suppression mechanism which might be
attributed to color screening in a QGP phase \cite{SatzQM99}.

\vspace{1cm} The authors acknowledge inspiring discussions with J.
Aichelin, S. A. Bass, G. E. Brown, C. Greiner, M. Gyulassy, C. M.
Ko, U. Mosel, R. Rapp, H. Satz, H. Sorge, H. St\"ocker, J.
Wambach, X.-N. Wang and K. Werner.

\newpage

\begin{figure}[t]
\phantom{a}\vspace*{-2.5cm}
\centerline{\psfig{figure=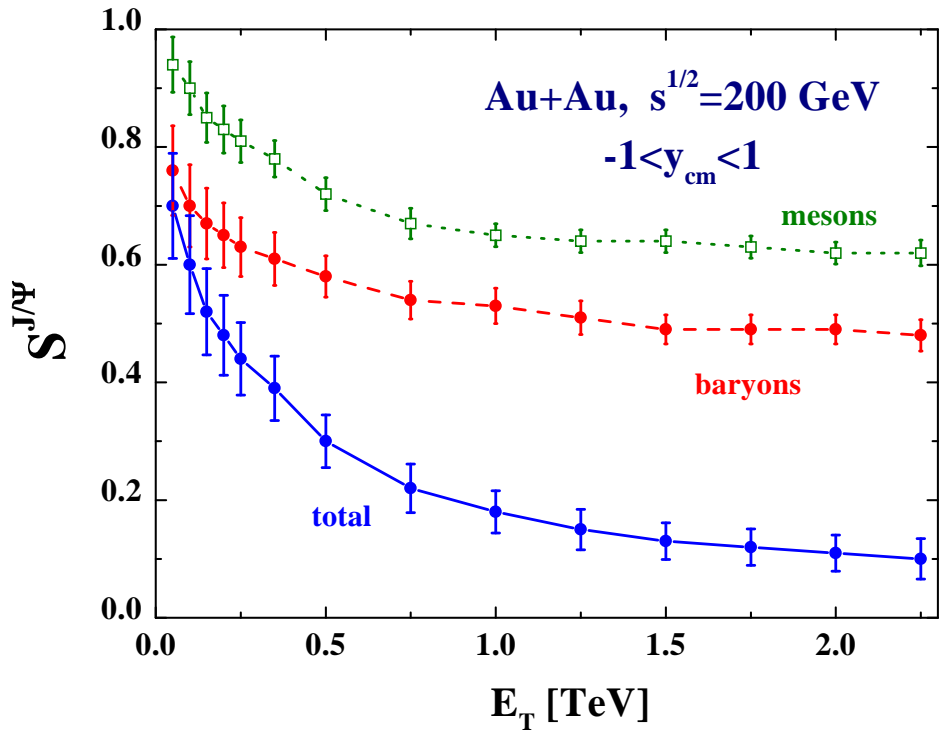,width=16cm}} \vspace*{-5cm}
\caption{The $J/\Psi$ survival factor $S_{J/\Psi}$ as a function
of the transverse energy $E_T$ in the rapidity interval (-1 $\leq
y_{cm} \leq$ 1) in Au + Au collisions at $\sqrt{s}$ = 200 GeV in
the 'late' comover model. The solid line represents the result for
$J/\Psi$ dissociation on nucleons and mesons, whereas the dashed
line and the dotted line correspond to the absorption on baryons
and mesons, respectively. The error bars in the figure are due to
statistics only. } \label{Fig14}
\end{figure}

\end{document}